\documentclass[aps,prx,twocolumn,nofootinbib]{revtex4-2}
\usepackage{amsmath}
\usepackage{amsfonts}
\usepackage{graphicx}
\usepackage{color}
\usepackage{dsfont}
\usepackage{verbatim}
\usepackage{mathtools}
\usepackage[normalem]{ulem}
\usepackage{comment}
\usepackage{stackrel}
\usepackage{bm}
\usepackage[pdftex]{hyperref}
\hypersetup{colorlinks = true, urlcolor = black, linkcolor = black, citecolor = black}

\DeclareMathOperator{\sgn}{sgn}

\definecolor{myblue}{rgb}{.93, .93, 1}

\newcommand{\bsub}{\begin{subequations}}
	\newcommand{\esub}{\end{subequations}}

\newcommand{\tr}{\mathsf{Tr}}

\begin{document}
	
	\title{Constrained motions and slow dynamics in one-dimensional bosons with double-well dispersion}
	
	\author{Yang-Zhi~Chou}\email{yzchou@umd.edu}
	\affiliation{Condensed Matter Theory Center and Joint Quantum Institute, Department of Physics, University of Maryland, College Park, Maryland 20742, USA}
	
	\author{Jay~D.~Sau}
	\affiliation{Condensed Matter Theory Center and Joint Quantum Institute, Department of Physics, University of Maryland, College Park, Maryland 20742, USA}
	
	\date{\today}

	\begin{abstract}
		We demonstrate slow dynamics and constrained motion of domain walls in one-dimensional (1D) interacting bosons with double-well dispersion. In the symmetry-broken regime, the domain-wall motion is ``fractonlike'' -- a single domain wall cannot move freely, while two nearby domain walls can move collectively. Consequently, we find an Ohmic-like linear response and a vanishing superfluid stiffness, which are atypical for a Bose condensate in a 1D translation invariant closed quantum system. Near Lifshitz quantum critical point, we obtain superfluid stiffness $\rho_s\sim T$ and sound velocity $v_s\sim T^{1/2}$, showing similar unconventional low-temperature slow dynamics to the symmetry-broken regime. Particularly, the superfluid stiffness suggests an order by disorder effect as $\rho_s$ increases with temperature. Our results pave the way for studying fractons in ultracold atom experiments.
	\end{abstract}
	
		\maketitle

\textit{Introduction.---} Ultracold neutral atom systems have been a promising platform for studying novel quantum many-body phenomena. Particularly, the ability to control interacting bosons motivates substantial new fundamental research \cite{Stenger1998spin,Greiner2002quantum,Kinoshita2004,Kinoshita2006,Sadler2006spontaneous,Lin2011spin,Zhang2012,Nguyen2014collisions,Beeler2013spin,Parker2013direct,JimenezGarcia2015,Clark2016,Putra2020,Yao2022domain} that does not have solid-state analogs. For example, interacting bosons with double-well dispersion (with two dispersion minima at $k=\pm k^*$) can be realized in the experiments \cite{Lin2011spin,Parker2013direct,Clark2016} with at least three distinct approaches --- One can achieve double-well dispersion by using two counterpropagating Raman laser lights that effectively create spin-orbit coupling for the pseudospin-$1/2$ bosons \cite{Lin2011spin,Galitski2013spin,Zhai2015review}. Alternatively, a bosonic ladder with $\pi$ flux per plaquette (by laser-assisted tunneling \cite{Goldman2014light}) generates double-well dispersion with the chain degrees of freedom acting like the pseudospins \cite{Atala2014observation,Celi2014,Dhar2012}.
Lastly, shaking an optical lattice with a frequency close to the energy difference between the ground band and the first excited band can realize double-well dispersion \cite{Parker2013direct,Clark2016}. Interacting bosons with double-well dispersion allow for rich quantum phase diagrams and novel dynamical response \cite{Ho2011,Li2012,Hu2012,Cole2012,Qu2013,Zheng2014,Xu2014,Khamehchi2016spin,Liu2016,Cole2019emergent,Orignac2017,Tokuno2014ground,Po2014,Po2015,Radic2015,Sur2019,Lake2021}.

Bose condensates with double-well dispersion are highly nontrivial, even without internal degrees of freedom (e.g., pseudospin). The two dispersion minima can be viewed as $\mathrm{Z}_2$ degrees of freedom, and a $\mathrm{Z}_2$ symmetry-breaking phase transition (analogous to an Ising ferromagnetic transition \cite{Lin2011spin}) occurs at low temperatures for repulsively interacting bosons. Topological defects appear as domain walls separating regimes with different momenta.
Intriguingly, the domain walls are stable and can persist for hundreds of milliseconds in the experiments \cite{Parker2013direct,Clark2016}, implying slow relaxation in the low-temperature (but $T\neq 0$) symmetry-broken regime.

	\begin{figure}[t]
	\includegraphics[width=0.325\textwidth]{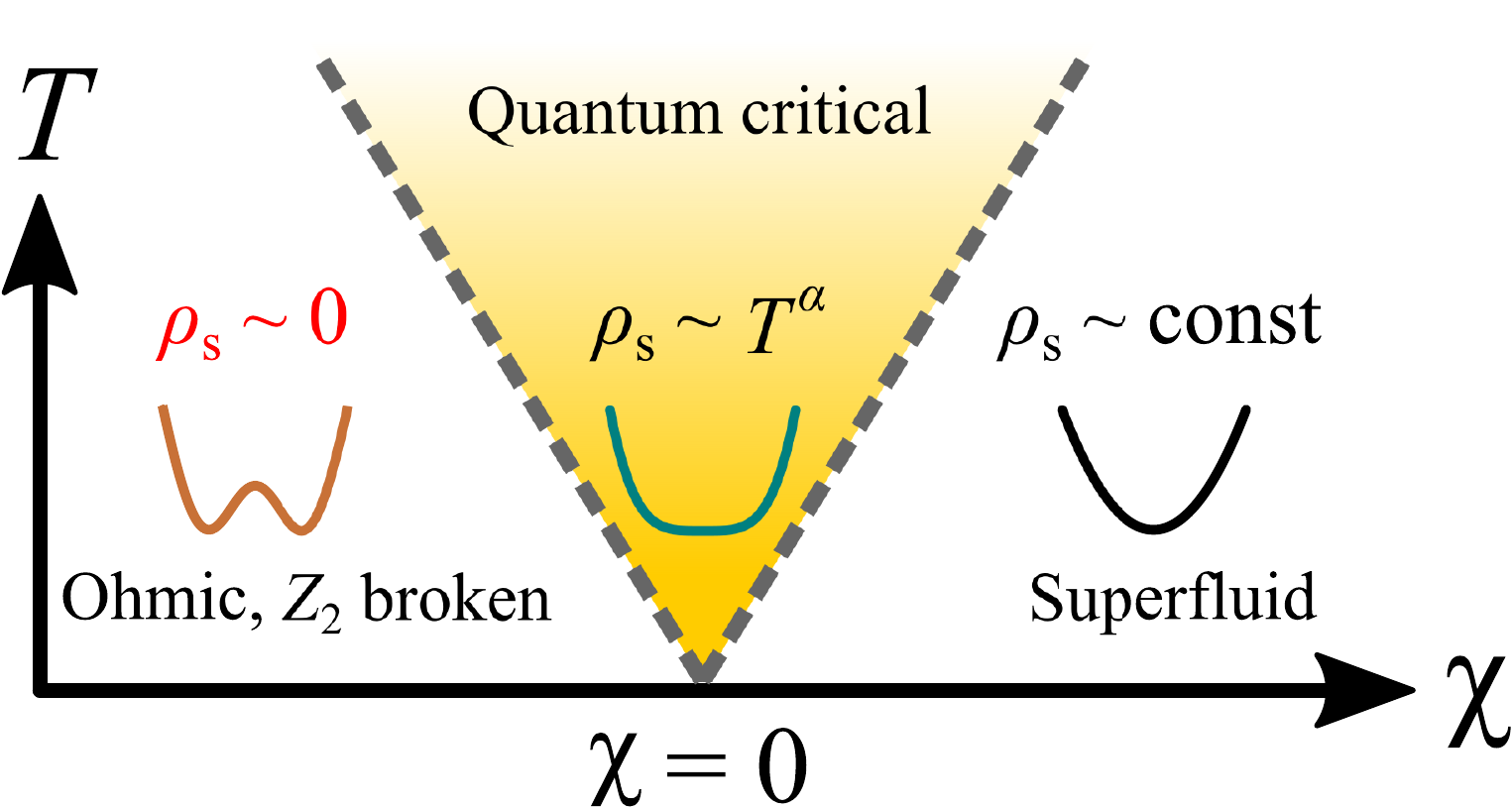}
	\caption{Phase diagram and superfluid stiffness ($\rho_s$). $\chi\propto-B$ is the control parameter of the quantum phase transition. For $\chi>0$, a dispersion with single minima is realized. $\rho_s$ is finite and essentially temperature-independent. For $\chi<0$, the dispersion develops two minima at $\pm k^*$, and a spontaneous $\mathrm{Z}_2$ breaking takes place. $\rho_s$ vanishes in this regime, and the corresponding transport is Ohmic-like. At $\chi=0$, a Lifshitz dispersion (i.e., a $k^4$ dispersion) manifests. The renormalization group flows suggest an interacting fixed point \cite{Yang2004,Sachdev1996} rather than a quantum Lifshitz Gaussian fixed point. The superfluid stiffness $\rho_s\sim T^{\alpha}$ with $\alpha=1$. 
	}
	\label{Fig:QCP}
\end{figure}

In this work, we study the dynamics of one-dimensional (1D) interacting single-component bosons with double-well dispersion as summarized in Fig.~\ref{Fig:QCP}. Under spontaneous $\mathrm{Z}_2$ symmetry breaking, the system naturally realizes multiple domains carrying finite momenta, $k^*$ or $-k^*$. We demonstrate that the motion of domain walls are highly constrained. A single domain wall cannot move, while two nearby domain walls can move in a collective fashion. Such intriguing kinetic properties are due to an \textit{emergent} dipole moment conservation, which suggests a genuine connection to the ``fractons''  \cite{Chamon2005,Castelnovo2012topological,Haah2011,Vijay2016,Pretko2017,Pretko2017Generalized,Prem2017,FractonsReview,Gromov2020,Pretko2020fracton,Radzihovsky2020,Gromov2022fracton,Seiberg2020field,Pai2020,Lake2022,Gorantla2022,Gorantla2022_2plus1,Zechmann2022fractonic,Radzihovsky2022,Lake2023}. 
The constrained domain-wall motion here is in contrast to the dynamics of domain walls in the transverse field Ising model \cite{Sachdev1997,Sachdev1999quantum} or holons and spions in 1D antiferromagnets \cite{Albuquerque2009}.
We also develop a linear response theory for a symmetry-broken state with multiple domain walls and show vanishing superfluid stiffness and Ohmic transport, despite being a Bose condensate. Near the interacting fixed point, we develop a hydrodynamic description and find superfluid stiffness and sound velocity vanish at zero temperature, showing the incipience of slow dynamics. Remarkably, the superfluid stiffness $\rho_s\sim T$, suggesting an order by (thermal-)disorder effect. Our theory provides a natural explanation for the stable domain walls in experiments \cite{Parker2013direct,Clark2016} and suggests an unprecedented way to study fractons in AMO systems.

\textit{Model.---} The 1D interacting single-component bosons with a double-well dispersion are described by
\begin{align}\label{Eq:H_DW_boson}
	\hat{H}=\int dx\left[-B|\partial_x b|^2+C|\partial_x^2b|^2-\mu|b|^2+\frac{U}{2}|b|^4\right]
\end{align}
where $b$ is the annihilation operator for a boson, $B$ and $C>0$ are the coefficients controlling single-particle dispersion, $\mu$ is the chemical potential, and $U>0$ denotes the repulsive short-range interaction. In this Letter, we focus mainly on the $B>0$ scenario, which admits a double-well dispersion with two minima at $k=\pm k^*=\pm\sqrt{B/(2C)}$ and an energy barrier $\epsilon_0=B^2/(4C)$ at $k=0$. $B=0$ is a critical point that realizes a Lifshitz dispersion (i.e., $k^4$). For $B<0$, the problem is qualitatively similar to the well-known repulsive Lieb-Liniger model \cite{LiebLiniger} (upto some dispersion correction).

In this work, we focus only on the superfluid phase [i.e., $U/\left(B n_0\right) \ll 1$], where $n_0$ is the density. Since there are two dispersion minima ($k=\pm k^*$), it is important to determine the ground state configuration. With mean-field approximation, one can show that the ground state is the same as the ``plane-wave phase'' in the 1D spin-orbit-coupled BEC \cite{Zhai2015review}, where only one minimum is occupied. As a result, the ground state features a spontaneous $Z_2$ symmetry breaking, and the ground state degeneracy is two.
We adopt the standard harmonic fluid approximation in the high-density superfluid limit \cite{Haldane1981} such that the complex boson field is decomposed into the density and phase fields as follows:
\begin{align}\label{Eq:rho_phi}
	b(x)\approx\sqrt{n_0+\delta n(x)}e^{i\phi(x)},
\end{align}
where $n_0$ is the density, $\delta n$ encodes the local fluctuation of density, and $\phi$ is the phase field. Using the expression of $b$ in Eq.~(\ref{Eq:rho_phi}), we can rewrite Eq.~(\ref{Eq:H_DW_boson}) with the two dynamical variables, $\phi$ and $\delta n$. For $|\delta n|\ll n_0$, we can integrate out $\delta n_0$ in the imaginary-time path integral and obtain a phase-only action. After rescaling of the parameters, we obtain an imaginary-time action $\mathcal{S}_{\text{eff}}$ given by \cite{SM}
\begin{align}\label{Eq:S_eff}
	\mathcal{S}_{\text{eff}}\approx&\int\!\! d\tau dx\!\left[\frac{1}{2}(\partial_{\tau}\theta)^2\!+\!\frac{1}{2}(\partial_x^2\theta)^2\!+\!\frac{r}{2}(\partial_x\theta)^2\!+\!u(\partial_x\theta)^4\!\right]\!,
\end{align}
where $\tau$ is the rescaled imaginary time, $\theta$ is the rescaled phase field, $r\propto -B$, and $u$ is the effective interaction of the phase fields. Equation~(\ref{Eq:S_eff}) is strictly valid for $\delta\equiv\mu/\epsilon_0\gg1$. For $\delta\ll 1$, density fluctuation cannot be ignored near a domain wall \cite{Liu2016}. We focus only on the limit $\delta \gg 1$ and $u>0$. Since much of our analysis ultimately relies on the low energy degrees of freedom, i.e., domain walls and phonons our conclusions are not qualitatively changed in the other limit as discussed in the Supplemental Material \cite{SM}.

\begin{figure}[t]
	\includegraphics[width=0.3\textwidth]{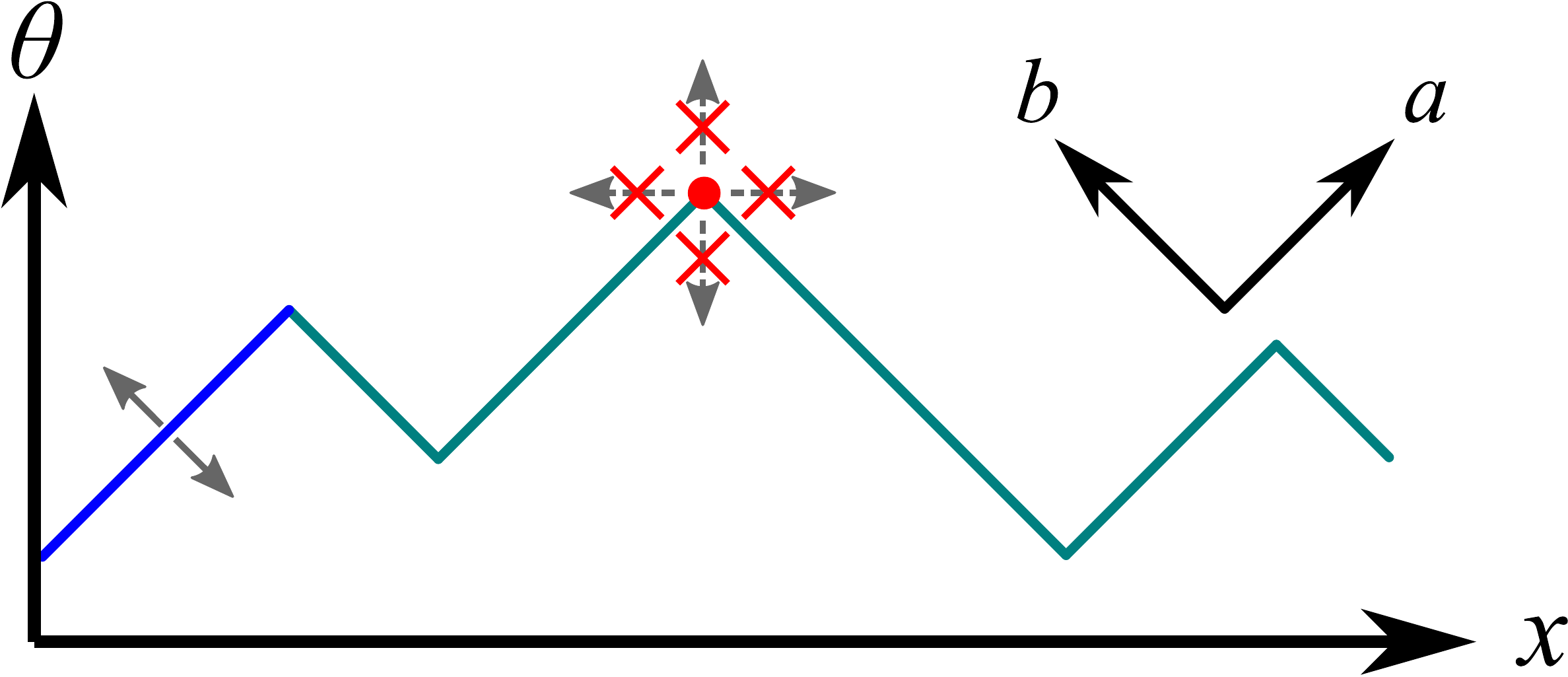}
	\caption{Motion of domain wall in the symmetry-broken phase. In each domain, $|\partial_x\theta|=m_0$, where $m_0=\sqrt{|r|/(4u)}$. A single domain wall (e.g., the red dot) cannot move freely because of the energy penalty, while an entire domain (e.g., the blue segment) can move. The directions of collective coordinates $a$ and $b$ correspond to the movement of domains. The domain-wall positions are labeled by $x_n$.}
	\label{Fig:DW_motion}
\end{figure}

\textit{Constrained motion and conservation of dipole moments.---} The 1D bosons with a double-well dispersion manifest spontaneous $Z_2$ symmetry breaking, analogous to a ferromagnetic transition. To see this, we introduce $m(x)=\partial_x\theta$, which corresponds to the momentum density of the superfluid. The static part of Eq.~(\ref{Eq:S_eff}) becomes the standard Landau theory for an Ising magnet, $\frac{r}{2}m^2+\frac{1}{2}(\partial_x m)^2+um^4$. For $r<0$, the $\langle m\rangle\neq 0$ features a spontaneous symmetry breaking. At zero temperature, the $m$ is spatially uniform, and $|m|=m_0=\sqrt{|r|/(4u)}$. At small finite temperatures, the system develops multiple domains with alternating signs of $m$ (corresponding to the slope of $\theta$) as illustrated in Fig.~\ref{Fig:DW_motion}. The density of domain walls is proportional to $\exp\left(-E_{\text{DW}}/T\right)$, where $E_{\text{DW}}$ is the energy cost for creating one domain wall \cite{Liu2016}. The dynamics in a state with multiple domain walls is highly unusual as we show in the following.

First, we discuss the single-domain-wall solution. An ``up-pointing'' single-domain-wall is described by \cite{Liu2016}
\begin{align}\label{Eq:DW_solution}
	\theta_{\text{DW}}(x)=\theta_0+m_0\sqrt{\frac{2}{|r|}}\ln\left[\cosh\left(\sqrt{\frac{|r|}{2}}(x-x_0)\right)\right],
\end{align}
where the domain-wall position is $x_0$. When $\sqrt{|r|}|x-x_0|\gg1$, $\theta_{\text{DW}}(x)$ recovers the slope $m_0$ for $x> x_0$ and $-m_0$ for $x< x_0$. 
Remarkably, moving a single domain wall will violate the energy constraint in Hamiltonian by forcing slopes to deviate from the equilibrium value $ \pm m_0 $. Thus, the motion of a single domain wall is suppressed due to the potential energy. However, one can move the entire domain while satisfying the potential energy (the blue segment in Fig.~\ref{Fig:DW_motion}). As a result, two nearby domain walls can move simultaneously. The constrained domain-wall motion here is a direct consequence of momentum conservation (i.e., spatial translation invariant) of the 1D interacting bosons with double-well dispersion because moving a single domain wall will result in change in the momentum of the condensate.

To understand the constrained domain-wall motion further, we examine the states with multiple domains more closely. First, we label the two types of domain walls to positive charge (up-pointing) and negative charge (down-pointing). The total dipole moment of the domain-wall charges is given by
\begin{align}
	\mathcal{D}=\sum_n\left(x_{2n}-x_{2n-1}\right),
\end{align}
where $x_n$ indicates the position of the $n$th domain wall (as illustrated in Fig.~\ref{Fig:DW_motion}). The alternating domains can be characterized by $\frac{\theta(x_{n+1})-\theta(x_{n})}{x_{n+1}-x_n}=(-1)^{n+1}m_0$ without loss of generality. Using this configuration, we can show that
\begin{align}
	\mathcal{D}=m_0^{-1}\sum_n\left[\theta(x_{2n})-\theta(x_{2n-1})\right]=2\pi Qm_0^{-1},
\end{align}
where $Q=\frac{1}{2\pi}\int dx\partial_x\theta$ is related to the total momentum, which governs the boundary condition. Thus, the total dipole moment $\mathcal{D}$ is a conserved quantity associated with the boundary condition of $\theta$. We note that the conservation of $\mathcal{D}$ (the dipole moment of topological defects) is dictated by the energy constraint, and the dipole moment conservation is an emergent low-temperature description when phonons can be ignored.
The conservation of dipole moment suggests a relation to the fractons \cite{Chamon2005,Castelnovo2012topological,Haah2011,Vijay2016,Pretko2017,Pretko2017Generalized,Prem2017,FractonsReview,Pretko2020fracton,Radzihovsky2020,Gromov2022fracton,Seiberg2020field,Pai2020,Lake2022,Gorantla2022,Gorantla2022_2plus1,Zechmann2022fractonic,Radzihovsky2022,Lake2023} that is known for its constrained dynamics of excitations. Our result suggests that the domain walls of 1D bosons with double-well dispersion can be viewed as fractons.

\textit{Phonon and relaxation mechanism.---} In addition to domain walls, the low energy dynamics of the system contains gapless phonon degrees of freedom as well. To understand the interplay between phonons and domain walls, we consider a long-wavelength variation $\delta\theta(x)$ on top of a single domain-wall profile $\theta_{\text{DW}}$ [Eq.~(\ref{Eq:DW_solution})]. We can construct a solution such that the entire $x<x_0$ domain displaces slightly (corresponding to the blue domain motion in Fig.~\ref{Fig:DW_motion}) while the $x>x_0$ domain remains the same. For $|x|\sqrt{|r|/2}\gg1$ (i.e., sufficiently away from the domain wall), we find that $\delta\theta(x\rightarrow-\infty)\neq0$ and $\delta\theta(x\rightarrow\infty)=0$, corresponding to a perfect reflection at the domain wall \cite{SM}. The phonons in each domain couple through the motion of the domain walls. Thus, we can integrate out the phonons in each domain wall and focus on the dynamics of the domain walls.

Integrating out the nearly perfectly reflecting phonons leads to two forces on the domain walls -- a Casimir effect and phonon drag. The Casimir effect is generated by the standing waves formed by the phonons in each domain, and it tends to stabilize configurations with equally spaced domain walls. The phonon drag is a friction force that arises from the ``radiation pressure'' as a moving domain wall experiences imbalance fluxes of momentum on the two sides (due to the longitudinal Doppler shift). The phonon drag can be described by a force $F_{\text{drag}}=-\gamma v$, where $\gamma$ is the coupling constant. The phonon fluctuations responsible for the drag also lead to diffusive motion of the domains with a velocity determined by the fluctuation dissipation theorem \cite{SM}. A direct consequence of the domain diffusion is an unusually slow dynamics (as compared to other systems, e.g., the transverse-field Ising model \cite{Sachdev1997,Sachdev1999quantum}). See \cite{SM} for a discussion.

\textit{Ohmic response and vanishing superfluid stiffness.---} To further quantify the slow dynamics of the domain walls, we study the transport properties in the symmetry-broken regime. Transport in the condensate is determined by the response to a vector potential $A\ge 0$, equivalent to tilting the optical lattice in the experiments \cite{Greiner2002quantum,Preiss2015strongly}. The vector potential $A$ and $\theta$ satisfy the following gauge transformation: $A\rightarrow A+\partial_x\Lambda$ and $\theta\rightarrow\theta+\Lambda$.
Therefore, we can incorporate the effect of vector potential by the minimal substitution: $\partial_x\theta\rightarrow\partial_x\theta-A$. In the presence of a uniform vector potential $A$, the minimal momenta become $m_0+A$ and $-m_0+A$, indicating that $ A$ modifies the slope in each domain. Assuming $0<A<m_0$, one can easily find new configurations that follow the change of slopes in $\theta$ without changing the boundary phase $\Delta\theta$. In addition, the ground-state energy with $n$ domain walls ($n>1$), $E_n[\theta(x)]$, does not depend on $A$, suggesting an emergent rank-two gauge symmetry, $E_n[\theta(x)]=E_n[\theta(x)-Ax]$ \cite{Gromov2020}. Intuitively, such properties imply the absence of response to a finite $A$, indicating a state with zero superfluid stiffness despite locally being a Bose condensate. In fact, the supercurrent  (i.e., distortion of slope) due to an application of a vector potential can relax by dissipating energy into the phonon drag. The result is a finite relaxation time for the current that is similar to the decay of current following a transient electric field in an Ohmic conductor.

To confirm the absence of superfluid stiffness, we develop a linear response theory for the symmetry-broken states and derive the Ohmic transport \cite{SM}. The goal is to derive the effective action of $A$ by integrating out the the domain-wall degrees of freedom. For simplicity, we assume a strong Casimir potential such that the domain walls are equally spaced and the domain size is $\bar{l}$. In the presence of $A$, we assume $\partial_x\theta=(-1)^{n+1}m_0+h(x)$ for $x_n<x<x_{n+1}$, where $h(x)$ is a response to the applied vector potential $A$. Then, we integrate out the fluctuations at the Gaussian level and derive an effective action for $A$ as follows:
\begin{align}
	\nonumber&\mathcal{S}_{A,\text{eff}}\bigg|_{k=0}
	\equiv\frac{\bar{l}}{\beta}\sum_{\omega_m}Q(\omega_m)\tilde{A}(-\omega_m)\tilde{A}(\omega_m).
\end{align}
The ac conductivity and superfluid stiffness can be obtained by $\sigma_{ac}(\omega)\propto \frac{i}{\omega}Q(\omega_m\rightarrow -i\omega-0^+)$ and $\rho_s\propto Q(\omega_m=0)$.
When $\gamma \neq 0$, we obtain an Ohmic response in the real part of low-frequency conductivity
\begin{align}
	\text{Re}\left[\sigma_{\text{ac}}(\omega)\right]\propto&\frac{16m_0|r|^2\gamma\left(8m_0^2|r|+\gamma^2\right)\bar{l}}{\left(8m_0|r|\gamma\right)^2+\left[\left(8m_0^2|r|+\gamma^2\right)\omega\bar{l}\right]^2}.
\end{align}
Moreover, the superfluid stiffness $\rho_s$ vanishes exactly, suggesting insulating behavior in a Bose condensate. Although the analytical results are derived with the equal-spaced domain wall assumption, the qualitative results remain the same for general situations as apparent from the numerical results discussed later.

\begin{figure}[t]
	\includegraphics[width=0.475\textwidth]{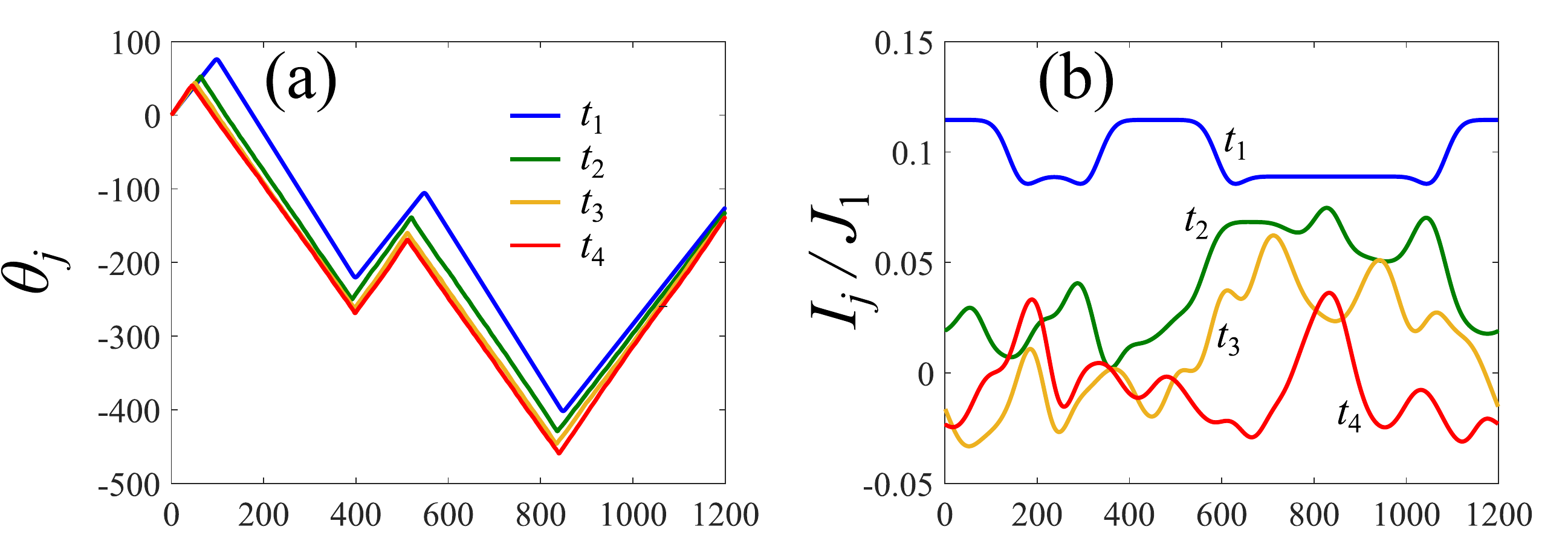}
	\caption{Numerical results for time evolution of phase and current profiles. An initial stationary state is prepared with a vector potential $A=\pi/30$ at $t=0$. Then, the state is evolved without a vector potential. (a) The phase configurations with different times. (b) The current configurations with different times. $J_1$ is the strength of the nearest-neighbor hopping in the lattice model. $t_1=0.0375J_1^{-1}$, $t_2=180J_1^{-1}$, $t_3=270J_1^{-1}$, and $t_4=360J_1^{-1}$. $L=1200$ for all the data. See Supplemental Material for a detailed discussion of the numerical procedures.}
	\label{Fig:GPE}
\end{figure}

In addition, we study the problem
using a discretized Gross-Pitaevski equation (GPE) \cite{SM}, which can simulate bosons in the semiclassical limit. 
The main goal of our simulation is to confirm the Ohmic response of the finite-temperature states with a few domain-walls.
To do this, we choose initial conditions $\psi_j=e^{i\theta_j}$ together with a choice for the phase-variable $\theta_j$ where the sign of the slope
of $\theta_j$ varies across domain walls in space. In addition, we assume that the system is subject to a large uniform electric field for a short
time, which as discussed in the previous subsection, corresponds to a tilting of the phase profile $\theta_j\rightarrow\theta_j+Aj$.
The ensuing dynamics obtained from the numerical solution of the GPE, shown in Fig.~\ref{Fig:GPE}(a), confirms the relaxation of the phase profile to a configuration
where the slopes obey the ground state value as time progresses through the simulation. 

To understand the observable transport consequences of this relaxation we compute the discrete local current operator. In Fig.~\ref{Fig:GPE}(b), we show the current profiles for a few representative times corresponding to the phase profiles in Fig.~\ref{Fig:GPE}(b). There are two important messages here. First, the current relaxes, suggesting a non-superfluid behavior. Second, the average current decreases substantially from the initial value, suggesting a vanishing current in the long-time limit. The decay of current confirms the Ohmic transport as predicted by our linear response theory.

In continuous 1D systems with momentum conservation, thermodynamic states can be associated with a certain momentum density. Such states, which result from the application of an electric field, carry a current even after the electric field is switched off. The resulting transport is effectively ballistic corresponding to infinite conductivity. In our case with $Z_2$ symmetry-broken ground states, the momentum imparted to the system can be absorbed into changing the configuration of the domain walls [see Fig.~\ref{Fig:GPE}]. Such a rearrangement transfers energy in the supercurrent into thermal energy of the phonons through a drag force on the domain walls. This dissipation of the current manifests as an Ohmic response of the current to an electric field. Our theory shows a rare example of zero superfluid stiffness and Ohmic response in a continuous translation invariant 1D system. In this case, the domain walls can be thought of as playing a similar role as the vortices in the high temperature phase of the two-dimensional superfluid where the Lorentz force on vortices from an applied supercurrent results in a dissipative voltage.

\textit{Lifshitz quantum hydrodynamics.--- } The slow dynamics of the symmetry-broken phase persists all the way to the vicinity of the Lifshitz quantum critical point \cite{JimenezGarcia2015,Clark2016,Cole2019emergent}. The quantum Lifshitz theory [i.e., Eq.~(\ref{Eq:S_eff}) with $r=u=0$] is at an unstable fixed point, and the renormalization group (RG) flows lead to an interacting fixed point with $r<0$ and $u>0$ \cite{Yang2004,Sachdev1996}. The scaling behavior in the vicinity of a quantum critical point can be analytically derived using RG and hydrodynamic treatment \cite{SM}. The main ideas and results are summarized in the following.

First, we construct a partition function incorporating the conservation laws (i.e., particle number, energy, and momentum). Based on the partition function, we derive the finite-temperature scalings of several observable quantities using the RG results. Particularly, $\rho_M\sim T^{-1}$ corresponds to diverging inertia at zero temperature. Concomitantly, the superfluid stiffness, $\rho_s\sim T$, vanishes at zero temperature \cite{SM}. The result of stiffness shows an order by thermal disorder effect as $\rho_s$ increases with temperature. Note that the classical gases with Lifshitz dispersion yield a different finite-temperature scaling in the inertia, $\rho_M\sim T^{-1/2}$ \cite{SM}. Another quantity of interest is the sound velocity, which can be derived using conservation laws and the thermodynamic relations. We find that the sound velocity $v_s\sim T^{1/2}$, which vanishes at zero temperature. We also note that the scaling of the Gaussian fixed point (i.e., $r=u=0$) yields the same results as discussed in Supplemental Material \cite{SM}. The vanishing of superfluid stiffness and sound velocity at low temperatures imply that the dynamics in the quantum critical regime is very slow, qualitatively similar to the symmetry-broken regime.

\textit{Discussion.---} The constrained dynamics due to the dipole moment conservation in the symmetry-broken regime indicates a connection to the fractons \cite{Chamon2005,Castelnovo2012topological,Haah2011,Vijay2016,Pretko2017,Pretko2017Generalized,Prem2017,Gromov2020,FractonsReview,Pretko2020fracton,Radzihovsky2020,Gromov2022fracton,Seiberg2020field,Pai2020,Lake2022,Gorantla2022,Gorantla2022_2plus1,Zechmann2022fractonic,Radzihovsky2022,Lake2023}. In addition, the conservation of dipole moment in our model is analogous but also distinct to the $S_z$ conservation in several spin-$1/2$ models \cite{Yang2020,Bastianello2022} that demonstrate Hilbert space fragmentation \cite{Khemani2020,Sala2020,Rakovszky2020,Yang2020,DeTomasi2019,Moudgalya2022,Kohlert2021experimental,Mukherjee2021,Bastianello2022,Ghosh2023prethermal}. Both conservation laws lead to slow dynamics -- {however, the dipole moment $\mathcal{D}$ in this case is not microscopic but rather associated with topological defects.
In contrast to systems with Hilbert space fragmentation, phonons together with slow domain motion will cause thermalization on an exponentially long timescale. This is similar to slow quantum relaxation due to dynamical constraints \cite{Lan2018}.
This long-time dynamics would include the effect of the Casimir force, which can also lead to an exponentially small in temperature residual superfluid stiffness.

The emergent dipole conservation in the symmetry-broken phase suggests that exact dipole conserving hydrodynamics \cite{Gorantla2022} with vanishing superfluid stiffness and associated slow dynamics of the $u=0$ Lifshitz critical point characterizes the critical point of our model. However, the finite $u>0$ is a relevant perturbation that results in a different quantum critical point \cite{Yang2004,Sachdev1996}. Despite this, the slow dynamics at the critical point \cite{Clark2016} are found to survive in the form of vanishing superfluid stiffness and sound velocity.
It is known that terms such as the $i\partial_{\tau}(\partial_x\theta)^2$ term that we ignore in our analysis can destabilize the quantum critical point in favor of a quantum fluctuation driven first order transition \cite{Kozii2017}. However, we expect our results to remain valid except very close to the quantum critical point.

Finally, we discuss the emergent symmetry in the low-energy symmetry-broken regime. The ground state energy with $n$ domain walls ($n>1$) does not depend on the spatially uniform vector potential $A$, implying an emergent rank-two gauge symmetry \cite{Gromov2020}. In addition to the vanishing superfluid stiffness, the emergent symmetry may be relevant to the several interesting features discussed in this Letter. Understanding the relation between this emergent symmetry and the slow dynamics in the symmetry-broken regime is an interesting future direction.

\begin{acknowledgments}
	We thank Maissam Barkeshli, Cheng Chin, Ian Spielman, Matthew Foster, Andrey Gromov, Han Pu, Chunlei Qu, Krishnendu Sengupta, and Zhi-Cheng Yang for useful discussions.
	This work is supported by the Laboratory for Physical Sciences, by JQI-NSF-PFC (Y.-Z.C.), by ARO W911NF2010232 (Y.-Z.C.), and by NSF DMR – 1555135 (J.D.S.).
\end{acknowledgments}

%%\bibliography{LFT}
%%%%%%%%%%%%%%%%%%%%

%%%%%%%%%%%%%%%%%%%%

\newpage \clearpage 

\onecolumngrid

\begin{center}
	{\large
		Constrained motions and slow dynamics in one-dimensional bosons with double-well dispersion
		\vspace{4pt}\\
		SUPPLEMENTAL MATERIAL
	}
\end{center}

\setcounter{figure}{0}
\renewcommand{\thefigure}{S\arabic{figure}}
\setcounter{equation}{0}
\renewcommand{\theequation}{S\arabic{equation}}

	In this supplemental material, we provide some technical details for the main results in the main text.

\section{Derivation of effective action}\label{App:Effective action}

We consider 1D complex bosons described by
\begin{align}\label{Eq:S_b}
	\mathcal{S}=\int\limits_{\tau,x}\left[b^*\partial_{\tau}b-B|\partial_x b|^2+C|\partial_x^2b|^2+U|b|^4-\mu|b|^2\right],
\end{align}
where $b$ is the complex boson field. We require $C>0$ for the stability of the theory. When $B<0$, the dispersion of the boson can be approximated by the conventional $k^2$ dispersion (upto a $k^4$ correction).
When $B>0$, the dispersion of the boson is a double-well. $B=0$ is the Lifshitz point.

In the high density limit, we use a density-phase representation for the boson field,
\begin{align}
	b(\tau,x)\approx \sqrt{n(\tau,x)}\,e^{i\phi(\tau,x)}.
\end{align}
We can express the derivatives of $b$ as follows:
\begin{align}
	b^*\partial_{\tau}b\rightarrow&\frac{1}{2}\partial_{\tau}n+n\left(i\partial_{\tau}\phi\right),\\
	\partial_xb\rightarrow&\frac{1}{2}\frac{(\partial_x n)}{n}b+b\left(i\partial_x\phi\right),\\
	\left|\partial_xb\right|^2\rightarrow&\frac{1}{4}\frac{\left(\partial_xn\right)^2}{n}+n\left(\partial_x\phi\right)^2,\\
	\partial_x^2b\rightarrow&\frac{1}{2}\frac{(\partial_x^2 n)}{n}b-\frac{1}{4}\frac{\left(\partial_x n\right)^2}{n^2}b-b\left(\partial_x\phi\right)^2+i\left[\frac{\partial_x n}{n}\left(\partial_x\phi\right)b+b\left(\partial_x^2\phi\right)\right]\\
	\nonumber\left|\partial_x^2b\right|^2\rightarrow&
	\frac{1}{4}\frac{\left(\partial_x^2n\right)^2}{n}+\frac{1}{16}\frac{\left(\partial_xn\right)^4}{n^3}+n\left(\partial_x\phi\right)^4-\frac{1}{4}\frac{\left(\partial_x^2n\right)\left(\partial_xn\right)^2}{n^2}-\left(\partial_x^2n\right)\left(\partial_x\phi\right)^2\\
	&+\frac{1}{2}\frac{\left(\partial_xn\right)^2}{n}\left(\partial_x\phi\right)^2+\frac{\left(\partial_xn\right)^2}{n}\left(\partial_x\phi\right)^2+n\left(\partial_x^2\phi\right)^2+2\left(\partial_x^2\phi\right)\left(\partial_x\phi\right)\left(\partial_xn\right)
\end{align}

The action given by Eq.~(\ref{Eq:S_b}) becomes to
\begin{align}
	\label{Eq:S_n_theta}\mathcal{S}\rightarrow \int\limits_{\tau,x}
	\left\{\begin{array}{c}
		\frac{1}{2}\partial_{\tau}n+n\left(i\partial_{\tau}\phi\right)-Bn\left(\partial_x\phi\right)^2+Un^2-\mu n\\[2mm]
		+C\left[n\left(\partial_x^2\phi\right)^2+n\left(\partial_x\phi\right)^4-\left(\partial_x^2n\right)\left(\partial_x\phi\right)^2+2\left(\partial_x^2\phi\right)\left(\partial_x\phi\right)\left(\partial_xn\right)\right]\\[2mm]
		-B\frac{1}{4}\frac{\left(\partial_xn\right)^2}{n}+C\left[\frac{1}{4}\frac{\left(\partial_x^2n\right)^2}{n}+\frac{1}{16}\frac{\left(\partial_xn\right)^4}{n^3}-\frac{1}{4}\frac{\left(\partial_x^2n\right)\left(\partial_xn\right)^2}{n^2}+\frac{3}{2}\frac{\left(\partial_xn\right)^2}{n}\left(\partial_x\phi\right)^2\right]
	\end{array}
	\right\}.
\end{align}

We assume that $n(\tau,x)=n_0+\delta n(\tau,x)$, where $n_0$ is the uniform background density and $\delta n$ is the fluctuation. By minimizing the free energy, one find that $n_0=\mu/(2U)$.
When $\delta n\ll n_0$, the action [given by Eq.~(\ref{Eq:S_n_theta})] becomes to
\begin{align}
	\mathcal{S}\rightarrow&\int\limits_{\tau,x}
	\left\{\begin{array}{c}
		\frac{1}{2}(\partial_{\tau}\delta n)+(n_0+\delta n)\left(i\partial_{\tau}\phi\right)-B(n_0+\delta n)\left(\partial_x\phi\right)^2+U(n_0+\delta n)^2-\mu (n_0+\delta n)\\[2mm]
		+C\left[(n_0+\delta n)\left(\partial_x^2\phi\right)^2+(n_0+\delta n)\left(\partial_x\phi\right)^4
		-\left(\partial_x^2 \delta n\right)\left(\partial_x\phi\right)^2+2\left(\partial_x^2\phi\right)\left(\partial_x\phi\right)\left(\partial_x \delta n\right)
		\right]\\[2mm]
		-B\frac{1}{4}\frac{\left(\partial_x \delta n\right)^2}{(n_0+\delta n)}+C\left[\frac{1}{4}\frac{\left(\partial_x^2 \delta n\right)^2}{(n_0+\delta n)}+\frac{1}{16}\frac{\left(\partial_x\delta n\right)^4}{(n_0+\delta n)^3}-\frac{1}{4}\frac{\left(\partial_x^2 \delta n\right)\left(\partial_x \delta n\right)^2}{(n_0+\delta n)^2}+\frac{3}{2}\frac{\left(\partial_x\delta n\right)^2}{(n_0+\delta n)}\left(\partial_x\phi\right)^2\right]
	\end{array}
	\right\}\\[2mm]
	=&\int\limits_{\tau,x}\left\{\begin{array}{c}
		\frac{1}{2}(\partial_{\tau}\delta n)+in_0(\partial_{\tau}\phi)\\[1mm]
		+U(\delta n)^2+2Un_0\delta n+Un_0^2-\mu \delta n-\mu n_0-B\frac{1}{4}\frac{\left(\partial_x \delta n\right)^2}{n_0}
		+C\frac{1}{4}\frac{\left(\partial_x^2\delta n\right)^2}{n_0}+\dots\\[2mm]
		-Bn_0(\partial_x\phi)^2+C n_0(\partial_x^2\phi)^2+C n_0(\partial_x\phi)^4\\[2mm]
		+i\delta n(\partial_{\tau}\phi)-B \delta n(\partial_x\phi)^2
		+C \delta n(\partial_x^2\phi)^2+C \delta n(\partial_x\phi)^4-2C\left(\partial_x^2 \delta n\right)\left(\partial_x\phi\right)^2+C\frac{3}{2}\frac{\left(\partial_x\delta n\right)^2}{n_0}\left(\partial_x\phi\right)^2+\dots
	\end{array}
	\right\}\\[2mm]
	=&\int\limits_{\tau,x}\left\{\begin{array}{c}
		+U\left[(\delta n)^2-\frac{B}{2\mu}\left(\partial_x \delta n\right)^2
		+\frac{C}{2\mu}\left(\partial_x^2\delta n\right)^2\right]+\dots\\[2mm]
		+\frac{\mu}{2U}\left[-B(\partial_x\phi)^2+C(\partial_x^2\phi)^2+C(\partial_x\phi)^4\right]\\[2mm]
		+i\delta n(\partial_{\tau}\phi)-B \delta n(\partial_x\phi)^2
		+C \delta n(\partial_x^2\phi)^2+C \delta n(\partial_x\phi)^4-2C\left(\partial_x^2 \delta n\right)\left(\partial_x\phi\right)^2
		+\frac{3CU}{\mu}\left(\partial_x\delta n\right)^2\left(\partial_x\phi\right)^2
		+\dots
	\end{array}
	\right\}.
\end{align}
In the last equation, $n_0=\mu/(2U)$ is used. The total derivative terms as well as the higher order in $\delta n/n_0$ terms are dropped.

The action $\mathcal{S}$ can be decomposed into three parts: $\mathcal{S}_n$ (density fluctuation), $\mathcal{S}_{\phi}$ (phase flucutaiton), and $\mathcal{S}_c$ (density-phase couplings). These actions are given by
\begin{align}
	\mathcal{S}_n=&U\int\limits_{\tau,x}\left[(\delta n)^2-\frac{B}{2\mu}\left(\partial_x \delta n\right)^2
	+\frac{C}{2\mu}\left(\partial_x^2\delta n\right)^2\right],\\
	\mathcal{S}_{\phi}=&\frac{\mu}{2U}\int\limits_{\tau,x}\left[-B(\partial_x\phi)^2+C(\partial_x^2\phi)^2+C(\partial_x\phi)^4\right],\\
	\mathcal{S}_c\approx&\int\limits_{\tau,x}\!\!\delta n\left[i(\partial_{\tau}\phi)\!-\!B (\partial_x\phi)^2
	\!+\!C (\partial_x^2\phi)^2\!+\!C (\partial_x\phi)^4
	\right]\!,
\end{align}
where we have dropped a few more irrelevant terms in $\mathcal{S}_c$. 
The density fluctuation is controlled by $U$ while the phase fluctuation is controlled by $n_0=\mu/(2U)$.

Formally, one can integrate out the density fluctuation at the Gaussian level and construct an effective action of the phase mode. The effective action is given by
	\begin{align}
		\mathcal{S}_{\text{eff}}=&\frac{\mu}{2U}\int\limits_{\tau,x}\left[-B(\partial_x\phi)^2+C(\partial_x^2\phi)^2+C(\partial_x\phi)^4\right]-\frac{1}{U}\int\limits_{\tau,x}\frac{\left[i(\partial_{\tau}\phi)-B(\partial_x\phi)^2
			+C(\partial_x^2\phi)^2+C(\partial_x\phi)^4\right]^2}{1+\frac{B}{2\mu}\partial_x^2+\frac{C}{2\mu}\partial_x^4}\\
		\approx&\frac{\mu}{2U}\int\limits_{\tau,x}\left[-B(\partial_x\phi)^2+C(\partial_x^2\phi)^2+C(\partial_x\phi)^4\right]
		+\int\limits_{\tau,x}\left[\frac{1}{U}\left(\partial_{\tau}\phi\right)^2+2i\frac{B}{U}\left(\partial_{\tau}\phi\right)\left(\partial_{x}\phi\right)^2
		+\frac{B}{2\mu U}\left(\partial_{\tau}\partial_x\phi\right)^2-\frac{B^2}{U}\left(\partial_x\phi\right)^4
		\right]\\
		\label{Eq:S_eff_bare}=&\frac{\mu}{2U}\int\limits_{\tau,x}\left[\frac{2}{\mu}(\partial_{\tau}\phi)^2+i\frac{4B}{\mu}(\partial_{\tau}\phi)(\partial_{x}\phi)^2
		+\frac{B}{\mu^2}(\partial_{\tau}\partial_x\phi)^2
		-B(\partial_x\phi)^2+C(\partial_x^2\phi)^2+\left(C-\frac{2B^2}{\mu}\right)(\partial_x\phi)^4\right].
	\end{align}
The above effective action has a few interesting properties. First of all, the $i(\partial_{\tau}\phi)(\partial_{x}\phi)^2$ appears. This term is believed to drive the phase transition to a first order transition. Otherwise, higher order terms are required. The $(\partial_{\tau}\partial_x\phi)^2$ corresponds to the dynamical term of the gauge field. The stability of the effective theory is set by $C-\frac{2B^2}{\mu}>0$, and the value of optimal momentum $k^*=\partial_x\phi$ becomes $\pm\sqrt{\frac{B}{2C-4B^2/\mu}}$, which recovers the noninteracting value $\pm\sqrt{\frac{B}{2C}}$ when $\mu\rightarrow\infty$. The renormalized value of $k^*$ is a manifestation of interaction effect.

To simplify the effective action, we introduce rescaling parameters, $\theta=\mathcal{A}^{-1}\phi$ and $\tilde{\tau}=\mathcal{B}^{-1}\tau$, where $\mathcal{A}=(\frac{1}{2\mu C})^{1/4}U^{1/2}$ and $\mathcal{B}=\left(\frac{2}{\mu C}\right)^{1/2}$. Thus, the effective action given by Eq.~(\ref{Eq:S_eff_bare}) becomes
\begin{align}
	\mathcal{S}_{\text{eff}}=\int\limits_{\tilde\tau,x}\Bigg[&\frac{1}{2}\left(\partial_{\tilde{\tau}}\theta\right)^2+i\zeta(\partial_{\tilde\tau}\theta)(\partial_{x}\theta)^2
	+\frac{g}{2}(\partial_{\tilde\tau}\partial_x\theta)^2
	+\frac{r}{2}(\partial_x\theta)^2+\frac{1}{2}(\partial_x^2\theta)^2+u(\partial_x\theta)^4
	\Bigg],
\end{align}
where $\zeta=\frac{2B\mathcal{A}^3}{U}$, $g=\frac{B\mathcal{A}^2}{\mu\mathcal{B} U}$, $r=-\frac{\mu B}{U}\mathcal{A}^2\mathcal{B}$, and $u=\frac{\mu}{2U}(C-2B^2/\mu)\mathcal{A}^4\mathcal{B}$ are the rescaled parameters. In the main text, we drop $\zeta$ and $g$ terms. We also make a notational substitution $\tilde{\tau}\rightarrow \tau$ for the simplicity of presentation.

\section{Scattering problem at a domain wall}\label{App:Scattering}

A single-domain-wall solution is described by
\begin{align}
	\theta_{\text{DW}}(x)=\theta_0+m_0\sqrt{\frac{2}{|r|}}\ln\left[\cosh\left(\sqrt{\frac{|r|}{2}}x\right)\right],
\end{align}
where we have set the domain-wall position at $x=0$. The (real-time) equation of motion is given by
\begin{align}\label{Eq:EOM_0}
	\partial_{t}\left(\partial_{t}\theta\right)-\partial_x\left[r\left(\partial_x\theta\right)+4u\left(\partial_x\theta\right)^3-\partial_x^3\theta\right]=0.
\end{align}
Now, we substitute $\theta$ by $\theta_{\text{DW}}(x)+\delta\theta(t,x)$ and rewrite the equation of motion as follows:
\begin{align}\label{Eq:EOM_new}
	\partial_{t}^2\delta\theta-\partial_x\left[\begin{array}{c}
		-|r|(\partial_x\delta\theta)+12u\left(\partial_x\theta_{\text{DW}}\right)^2\left(\partial_x\delta\theta\right)\\
		+12u\left(\partial_x\theta_{\text{DW}}\right)\left(\partial_x\delta\theta\right)^2\\
		+4u\left(\partial_x\delta\theta\right)^3-(\partial_x^3\delta\theta)
	\end{array}
	\right]=0,
\end{align}
where we have used the fact that $\theta_{\text{DW}}$ obeys the equation of motion. For $\sqrt{|r|/2}|x|\gg 1$, we can derive the linearized equation of motion (ignoring $O(\delta\theta^2)$) as follows
\begin{align}	
	\nonumber&\partial_{t}^2\delta\theta-\partial_x\left[-|r|(\partial_x\delta\theta)+12um_0^2\left(\partial_x\delta\theta\right)-(\partial_x^3\delta\theta)\right]=0\\
	\label{Eq:LW_EOM}\rightarrow& \left[\partial_{t}^2-2|r|\partial_x^2+\partial_x^4\right]\delta\theta=0,
\end{align}
where we have used $m_0^2=|r|/(4u)$. The above wave equation gives the dispersion of the phonon in the linearized regime. In the long-wavelength limit, we obtain $\omega^2=v_p^2k^2$, where $v_p=\sqrt{2|r|}$.

The general scattering problem can be solved by studying Eq.~(\ref{Eq:EOM_new}). Here, we focus only on the long-wavelength limit. 
Since the problem has a reflection symmetry about $x=0$, we can rewrite $\delta\theta$ as $\delta\theta=a_S\theta_S+a_A\theta_A$, where
\begin{align}
	\theta_S(x)|_{|x|\rightarrow\infty}=&\cos\left(k|x|+\alpha_{S,k}\right),\\
	\theta_A(x)|_{|x|\rightarrow\infty}=&\cos\left(k|x|+\alpha_{A,k}\right)\sgn(x).
\end{align}
In the above expression, $\alpha_{S,k}$ and $\alpha_{A,k}$ are the phase factors. We can show that there are two zero modes ($\omega\rightarrow 0$) of the equation of motion in Eq.~(\ref{Eq:EOM_0}): (a) An overall constant shift in $\theta$ and (b) domain-wall translation. 
Thus, one can easily show that $\alpha_{S,k}=0$ and $\alpha_{A,k}=0$ for $k\rightarrow 0$. Then, we consider the conventional scattering waves far away from the domain wall ($x=0$), 
\begin{align}
	\delta\theta(x)|_{x\rightarrow-\infty}\propto &e^{ikx}+a_Re^{-ikx},\\
	\delta\theta(x)|_{x\rightarrow \infty}\propto &a_te^{ikx}.
\end{align}
We can easily show find that $a_S=-a_A$ satisfy the scattering ansatz with $a_r=1$ and $a_t=0$, corresponding to the perfect reflection at the domain wall. Therefore, we expect that the long-wavelength phonons form standing waves in each domain.

\section{Casimir effect and phonon drag} 

To study the Casimir effect, we compute the ``vacuum energy'' due to the standing waves of long-wavelength phonons in a domain with length $l$, corresponding to
\begin{align}
	F_{l}=\frac{1}{\beta}\sum_{m=1}^{m_c}\ln\left(1-e^{-\beta\omega_{k_m}}\right),
\end{align}
where $\beta$ is the inverse temperature, $\omega_{k}=v_pk$ denotes the phonon frequency in the long-wavelength limit, $v_p$ is the phonon velocity, $k_m=\pi m/l$ for the standing waves, and $m_c$ is the number of long-wavelength phonons. In the low-temperature limit, the Casimir potential is approximated by:
\begin{align}
	V_{\text{Casimir}}\approx-T\mathcal{C}\sum_n\ln\left[\frac{T}{\pi}(x_{n+1}-x_n)\right]+\text{const},
\end{align}
where $\mathcal{C}=m_c$.
The $V_{\text{Casimir}}$ with a fixed number of domain walls is minimized when the domain walls are equal-spaced, i.e., domain sizes are the same. Using the equipartition theorem, we obtain that the root-mean-square of domain-wall displacement is $\bar{l}/\sqrt{2\mathcal{C}}$, where $\bar{l}$ is the averaged domain size. For sufficiently low (but nonzero) temperatures, $\mathcal{C}$ is typically a large number. Thus, the domain wall fluctuation can be ignored. 

Let us now discuss the origin of phonon drag which is a friction force that arises from the motion of the domain walls relative to the phonons. Since a domain wall strongly scatter off the long-wavelength phonons, the domain wall is also subject to a ``radiation pressure'' related to the flux of momentum $v_p\mathcal{E}$, where $\mathcal{E}$ is the energy density of the phonons. The radiation pressure is balanced between the two sides at zero domain-wall velocity. At a finite velocity $v$ of a domain wall, the longitudinal Doppler shift gives $\mathcal{E}'(v)=\mathcal{E}\sqrt{\frac{1-v/v_p}{1+v/v_p}}$ for phonons move in the same direction as the domain wall. Thus, the phonons provide a drag force on the domain wall, described by 
\begin{align}\label{Eq:drag_F}
	F_{\text{drag}}=v_p\left[\mathcal{E}(v)-\mathcal{E}(-v)\right]=-\gamma v, 
\end{align}
where $\gamma\approx 2\mathcal{E}$ is the coupling constant of the phonon drag. The phonon drag causes a diffusive motion of the domain walls. Since single domain wall motion is forbidden, the dominant relaxation is the domain diffusion (i.e., a collective diffusion between two nearby domain walls). One can easily check that the domain-wall motion is also diffusive, but the two nearby domain walls are correlated. 

We briefly discuss the derivation of domain diffusion in the following. The 1D Langevin equation for a domain with length $l$ is given by
\begin{align}
	m\frac{d v}{dt}=-\gamma v+\delta F(t),
\end{align}
where $m=cl$ is the mass of the domain, $c$ is a constant, $\gamma$ is the frictional coefficient, and $\delta F(t)$ is the fluctuating force (i.e., noise). We consider $\langle\delta F(t)\rangle=0$ and $\langle\delta F(t) \delta F(t')\rangle=2\gamma T\delta(t-t')$, which are consistent with the fluctuation-dissipation theorem. The solution of Langevin equation is given by
\begin{align}
	v(t)=e^{-\gamma t/m}v(0)+\int_{0}^{t}ds e^{-\gamma(t-s)/m}\delta F(s)/m.
\end{align}
For $t\gg m/\gamma$, the first term of the velocity is completely suppressed, and $v(t)$ is dominated by the fluctuation. However, $m/\gamma\propto l\propto e^{\Delta/T}$ (where $\Delta$ is the energy cost of a domain wall), suggesting that the correlation time diverges in the low-temperature limit.
One can also show that $\lim\limits_{t\rightarrow\infty}\langle x^2(t)\rangle=2Dt$, where $D=T/\gamma$, independent of $l$.

\section{Estimate of autocorrelation time}\label{App:AutoT}

To understand the autocorrelation time in the interacting bosons with double-well dispersion, we first review several results in the transverse-field Ising model (TFIM) \cite{Sachdev1999quantum}. Then, we generalize the ideas for the momentum-momentum correlation function of interacting bosons with double-well dispersion.

The equal-time spin-spin correlation function of TFIM in the low-temperature symmetry-broken regime can be computed semicalssically. For a system $L\gg |x|$ with $N$ thermally excited domain walls, the density is $\rho=N/L=1/l_0$. The probability of finding a particle between $0$ and $x$ is given by $p=|x|/L$. It is crucial to note that each domain wall flips the sign of spin. With these ideas in mind, we can compute the expectation value of finding two equal-sign spins with a separation $|x|$ as follows:
\begin{align}
	C(x,0)\propto &\sum_{j=0}^N(-1)^j p^j(1-p)^{N-j}\frac{N!}{j!(N-j)!}\\
	=&(1-2p)^N=\left(1-\frac{2|x|}{L}\right)^{L\rho}
	\rightarrow e^{-|x|/\xi_c},
\end{align} 
where $\xi_c=1/(2\rho)$ dictates the length scale over which the spin correlation is missing. For TFIM, $\xi_c^{-1}=\sqrt{\frac{2\Delta T}{\pi c^2}}e^{-\Delta/T}$, $c$ is the velocity of excitation (domain wall) in TFIM, and $\Delta$ is the energy cost of creating an excitation (domain wall)\cite{Sachdev1997,Sachdev1999quantum}
One can also compute the general space-time correlation function $C(x,t)$ using the same idea and averaging the ballistic propagation of domain walls \cite{Sachdev1997,Sachdev1999quantum}. The idea is to estimate the time duration that a domain wall at $x=\xi_c$ travels to $x=0$. In the end, one derive the equal-space autocorrelation function $C(0,t)\propto e^{-|t|/\tau_{\text{TFIM}}}$, where $\tau_{\text{TFIM}}^{-1}=\frac{2}{\pi}Te^{-\Delta/T}$ \cite{Sachdev1997,Sachdev1999quantum}.

A simple way to estimate autocorrelation time is to use $\bar\tau_{\text{TFIM}}=\xi_c/v_T$, where $v_T$ is the thermal velocity obtained from the equipartition theorem. Since the dispersion of domain wall in TFIM is given by $\epsilon_k\approx \Delta+\frac{c^2k^2}{2\Delta}$, we obtain $v_T=c\sqrt{T/\Delta}$. Then, we obtain $\bar\tau_{\text{TFIM}}^{-1}=\sqrt{\frac{2}{\pi}}Te^{-\Delta/T}$, which differs from $\tau_{\text{TFIM}}^{-1}$ by an overall numerical prefactor. Thus, this simple analysis seems to provide a reasonable estimate of the autocorrelation time.

Now, we discuss the interacting bosons with double-well dispersion. The momentum $\partial_x\theta$ here is analogous to the spin in TFIM. Using the same idea for TFIM, the equal-time correlation function $\langle\partial_x\theta(x,t)\partial_x\theta(0,t)\rangle\propto \exp(-|x|/\xi_c)$, where $\xi_c=l_0/2$. The autocorrelation function is significantly different because the constrained domain-wall motion and the phonon drag. The phonon drag induces a scattering time $\tau_{sc}=m/\gamma$, where $m\sim l_0$ is the mass of a moving domain. Since the scattering due to phonon drag is at random, we assume a random-walk process for the domain motion. Each scattering gives a move with a distance $\delta X\sim v_T'\tau_{sc}$, where $v_T'\sim\sqrt{T/l_0}$ is obtained from equipartition theorem.
For $M$ times of scattering, the total displacement due to random walk is given by $\delta X\sqrt{M}$. Using $M=\tau_{DW}/\tau_{sc}$ (with $\tau_{\text{DW}}$ being the autocorrelation time) and $\delta X\sqrt{M}=\xi_c$, we obtain $\tau_{\text{DW}}\sim \gamma l_0^2/T$ which is significantly larger than $\tau_{\text{TFIM}}$ in the low-temperature limit.
For completeness, we discuss $\gamma=0$ case. The analysis is similar to the autocorrelation time for TFIM except that the thermal velocity $v_T'\sim \sqrt{T/l_0}$. Therefore, we obtain $\tau_{\text{DW}}'\sim l_0^{3/2}/\sqrt{T}$, which is intermediate between $\tau_{\text{TFIM}}$ and $\tau_{\text{DW}}$.

\section{Classical model for domain wall dynamics}\label{App:Classical_mapping}

In this section, we present a framework for studying the response to a vector potential in a symmetry-broken state with multiple domains. We first introduce the collective variables and the Lagrangian coordinates. Then, we develop a linear-response theory and compute the response function.

\subsection{Collective coordinate}

To describe the motion of domains, we introduce a set of collective variables as follows:
\begin{align}
	a(x)=\frac{1}{\sqrt{2}}\left[\theta(x)+m_0x\right],\,\,\,b(x)=\frac{1}{\sqrt{2}}\left[\theta(x)-m_0x\right].
\end{align}
Conversely,
\begin{align}
	\theta(x)=\frac{1}{\sqrt{2}}\left[a(x)+b(x)\right],\,\,\,x=\frac{1}{\sqrt{2}m_0}\left[a(x)-b(x)\right].
\end{align}
We can express $(x,\theta(x))$ in terms of $(a,b)$.

\subsection{General formalism for domain-wall dynamics}

We construct a general formalism for studying the domain wall dynamics in this subsection. To keep track of the domain walls, we introduce a Lagrangian coordinate $s$, and the Eulerian coordinate $x\equiv X(\tau,s)$. Formally, we can also view this as a reparametrization from $(\tau,x)$ to $(\tau,s)$. We require that $s_n$ labels the $n$th domain with position $X(\tau,s_n)$. Without loss of generality, we set $s_n=n$. Note that $x$ without specifying $s$ is just a variable, i.e., independent of $\tau$. The $\tau$-dependence is acquired when the value of $s$ is assigned. The total derivative of $X$ is expressed by
\begin{align}
	dx=\left(\frac{\partial X}{\partial \tau}\right)_{s}d\tau+\left(\frac{\partial X}{\partial s}\right)_{\tau}ds,
\end{align}
where the subscript $\tau$ ($s$) in the partial derivative means fixing $\tau$ ($s$). We derive two useful identities
\begin{align}
	\frac{dx}{dx}=1=\left(\frac{\partial X}{\partial \tau}\right)_{s}\frac{d\tau}{dx}+\left(\frac{\partial X}{\partial s}\right)_{\tau}\frac{ds}{dx}\rightarrow& dx=\left(\frac{\partial X}{\partial s}\right)_{\tau} ds\\
	\frac{dx}{d\tau}=0=\left(\frac{\partial X}{\partial \tau}\right)_{s}+\left(\frac{\partial X}{\partial s}\right)_{\tau}\frac{ds}{d\tau}\rightarrow&\frac{ds}{d\tau}=-\frac{\left(\frac{\partial X}{\partial \tau}\right)_{s}}{\left(\frac{\partial X}{\partial s}\right)_{\tau}}
\end{align}
We also define $\theta\equiv\tilde{\theta}(\tau,s)$. The derivatives of $\theta$ are given by
\begin{align}
	\partial_x\theta=&\frac{\left(\frac{\partial \tilde{\theta}}{\partial s}\right)_{\tau}}{\left(\frac{\partial X}{\partial s}\right)_{\tau}},\\
	\partial_{\tau}\theta=&\left(\frac{\partial \tilde{\theta}}{\partial \tau}\right)_s+\left(\frac{\partial \tilde{\theta}}{\partial s}\right)_\tau\frac{ds}{d\tau},
\end{align}
We consider a state such that
\begin{subequations}\label{Eq:tilde_theta_x}
	\begin{align}
		\tilde{\theta}(\tau,s)=&(s-n)\theta_{n+1}+(n+1-s)\theta_n,\\
		X(\tau,s)=&(s-n)x_{n+1}+(n+1-s)x_n,
	\end{align}
\end{subequations}
for $n\le s\le n+1$. In the above expression, $x_n\equiv X|_{s=n}$ and $\theta_n\equiv\tilde{\theta}|_{s=n}$ correspond to position and the phase variable of the $n$th domain wall. We note that Eq.~(\ref{Eq:tilde_theta_x}) does not incorporate the accurate shape of domain walls. Based on Eq.~(\ref{Eq:tilde_theta_x}), we compute
\begin{align}
	\partial_x\theta=&\frac{\theta_{n+1}-\theta_n}{x_{n+1}-x_{n}},\\
	\nonumber\partial_{\tau}\theta=&(s-n)\left[\partial_{\tau}\theta_{n+1}-(\partial_{\tau}x_{n+1})(\partial_x\theta)\right]\\
	&-(n+1-s)\left[\partial_{\tau}\theta_{n}-(\partial_{\tau}x_n)(\partial_x\theta)\right],
\end{align}
for $n<s<n+1$. 

Our goal is to study the symmetry-broken state with multiple domains. In the imaginary-time formalism, we can treat the problem as a classical string described by an effective Hamiltonian $H_{\text{eff}}=K+V$, where
	\begin{align}
		\nonumber K=&\int\limits_x\frac{1}{2}(\partial_{\tau}\theta)^2=\frac{1}{2}\sum_n\int_{n}^{n+1} ds(x_{n+1}-x_n)\left\{\begin{array}{c}
			(s-n)^2\left[\partial_{\tau}\theta_{n+1}-(\partial_{\tau}x_{n+1})(\partial_x\theta)\right]^2\\[1mm]
			+2(s-n)(n+1-s)\left[\partial_{\tau}\theta_{n+1}-(\partial_{\tau}x_{n+1})(\partial_x\theta)\right]\left[\partial_{\tau}\theta_{n}-(\partial_{\tau}x_n)(\partial_x\theta)\right]\\[1mm]
			+(n+1-s)^2\left[\partial_{\tau}\theta_{n}-(\partial_{\tau}x_n)(\partial_x\theta)\right]^2
		\end{array}\right\}\\
		=&\frac{1}{6}\sum_n(x_{n+1}-x_n)\left\{\begin{array}{c}
			\left[\partial_{\tau}\theta_{n+1}-(\partial_{\tau}x_{n+1})(\partial_x\theta)\right]^2\\[1mm]
			+\left[\partial_{\tau}\theta_{n+1}-(\partial_{\tau}x_{n+1})(\partial_x\theta)\right]\left[\partial_{\tau}\theta_{n}-(\partial_{\tau}x_n)(\partial_x\theta)\right]\\[1mm]
			+\left[\partial_{\tau}\theta_{n}-(\partial_{\tau}x_n)(\partial_x\theta)\right]^2
		\end{array}\right\},\\[2mm]
		V=&\int\limits_x\left[-\frac{|r|}{2}(\partial_x\theta)^2+\frac{1}{2}(\partial_x^2\theta)^2+\frac{|r|}{4m_0^2}(\partial_x\theta)^4\right]=\sum_n\int_{n}^{n+1}ds(x_{n+1}-x_n)\left\{\frac{|r|}{4m_0^2}\left[(\partial_x\theta)^2-m_0^2\right]^2+\frac{1}{2}(\partial_x^2\theta)^2\right\}.
	\end{align}
To derive the zero-point kinetic energy, we assume that $\partial_x\theta=(-1)^{n+1}m_0$ for $n<s<n+1$. Moreover, we ignore the $(\partial_x^2\theta)^2$ term in $V$ as it give rise to the domain wall energy, which we assume to be a constant for our fixed domain wall number calculations. Thus, the zero-point kinetic energy is expressed by
	\begin{align}
		\nonumber K_0=&\frac{1}{6}\sum_n(x_{n+1}-x_n)\left\{\begin{array}{c}
			\left[\partial_{\tau}\theta_{n+1}-(\partial_{\tau}x_{n+1})(-1)^{n+1}m_0\right]^2\\[1mm]
			+\left[\partial_{\tau}\theta_{n+1}-(\partial_{\tau}x_{n+1})(-1)^{n+1}m_0\right]\left[\partial_{\tau}\theta_{n}-(\partial_{\tau}x_n)(-1)^{n+1}m_0\right]\\[1mm]
			+\left[\partial_{\tau}\theta_{n}-(\partial_{\tau}x_n)(-1)^{n+1}m_0\right]^2
		\end{array}\right\}\\
		=&\frac{1}{3}\sum_n\left[(x_{2n+1}-x_{2n})\left(\dot{a}_{2n+1}^2+\dot{a}_{2n+1}\dot{a}_{2n}+\dot{a}_{2n}^2\right)+(x_{2n}-x_{2n-1})\left(\dot{b}_{2n}^2+\dot{b}_{2n}\dot{b}_{2n-1}+\dot{b}_{2n-1}^2\right)\right].
	\end{align}

	\subsection{Dissipative action}
	
	As we discuss in main text, an important momentum relaxation mechanism is through the scattering between domain walls and phonons give rise to a phonon drag. Here, we describe the dissipative action in the Lagrangian coordinate. With the standard treatment \cite{Kamenev2009keldysh}, the dissipative action (corresponding to phonon drag force $F=-\gamma v$) is given by
	\begin{align}
		\mathcal{S}_{\text{dis}}=&\frac{\gamma}{2}\frac{1}{\beta}\sum_{\omega_m}\sum_n|\omega_m|x_n(-\omega_m)x_n(\omega_m)
		=\frac{\gamma}{4m_0\beta}\sum_{\omega_m}\sum_n|\omega_m|\left[a_n(-\omega_m)-b_n(-\omega_m)\right]\left[a_n(\omega_m)-b_n(\omega_m)\right].
	\end{align}
	This dissipative action is crucial in the linear-response calculations as it generates an Ohmic response.

	\subsection{Vector potential and response}
	
	In this section, we discuss the derivation of linear-response in the presence of a small vector potential $A$.
	Our goal is to derive an effective action of $A$ by integrating out all the fluctuating domain-wall degrees of freedom at the Gaussian level. The quadratic coefficient of $A^2$ term in the effective action corresponds to the linear-response coefficient.
	
	In the presence of a vector potential, we can perform the minimal substitution: $\partial_x\theta\rightarrow \partial_x\theta-A$. The $V$ term becomes
	\begin{align}
		V=&\sum_n\int_{n}^{n+1}ds(x_{n+1}-x_n)\left\{\frac{|r|}{4m_0^2}\left[(\partial_x\theta-A)^2-m_0^2\right]^2+\frac{1}{2}(\partial_x^2\theta-\partial_xA)^2\right\}\\
		=&\sum_n\int_{n}^{n+1}ds(x_{n+1}-x_n)\left\{\begin{array}{c}
			\frac{|r|}{4m_0^2}\left[(\partial_x\theta)^2-m_0^2\right]^2+\frac{1}{2}(\partial_x^2\theta)^2\\[2mm]
			-A\frac{|r|}{m_0^2}(\partial_x\theta)\left[(\partial_x\theta)^2-m_0^2\right]-(\partial_xA)(\partial_x^2\theta)\\[2mm]
			+A^2\frac{|r|}{2m_0^2}\left[3(\partial_x\theta)^2-m_0^2\right]+\frac{1}{2}\left(\partial_xA\right)^2
		\end{array}
		\right\}+O(A^3),
	\end{align}
	where we keep upto $A^2$ order in the spirit of linear response. To incorporate the response to the vector potential, we assume that $\partial_x\theta=(-1)^{n+1}m_0+h(x)$ for $x_n<x<x_{n+1}$, where $h(x)$ is a response due to $A$ and is given by
	\begin{align}\label{Eq:h_x}
		h(x)=&\partial_x\theta-(-1)^{n+1}m_0
		=\frac{\theta_{n+1}-\theta_n-(-1)^{n+1}m_0(x_{n+1}-x_n)}{x_{n+1}-x_n}
		=\frac{\sqrt{2}}{x_{n+1}-x_n}\times\begin{cases}
			(a_{n+1}-a_{n}), \text{ for }n\text{ is even },\\
			(b_{n+1}-b_{n}), \text{ for }n\text{ is odd }.
		\end{cases}
	\end{align}
	
	Thus, we expand $V$ upto $O(h^2)$ as follows:
	
	\begin{align}
		V\approx&\sum_n\int_{n}^{n+1}ds(x_{n+1}-x_n)\left\{\begin{array}{c}
			\frac{|r|}{4m_0^2}\left[2(-1)^{n+1}m_0h+h^2\right]^2+\frac{1}{2}(\partial_x h)^2\\[2mm]
			-A\frac{|r|}{m_0^2}((-1)^{n+1}m_0+h)\left[2(-1)^{n+1}m_0h+h^2\right]-(\partial_xA)(\partial_xh)\\[2mm]
			+A^2\frac{|r|}{2m_0^2}\left[2m_0^2+6(-1)^{n+1}m_0h+3h^2\right]+\frac{1}{2}\left(\partial_xA\right)^2
		\end{array}
		\right\}\\
		=&\sum_n\int_{n}^{n+1}ds(x_{n+1}-x_n)\left\{\begin{array}{c}
			|r|h^2+\frac{1}{2}(\partial_x h)^2\\[2mm]
			-A\frac{|r|}{m_0^2}\left(2m_0^2h\right)-(\partial_xA)(\partial_xh)\\[2mm]
			+A^2\frac{|r|}{2m_0^2}\left[2m_0^2\right]+\frac{1}{2}\left(\partial_xA\right)^2
		\end{array}
		\right\}+O(h^3,Ah^2,A^2h)
	\end{align}
	We further ignore the terms involving $\partial_x h$ as they are not important for our purpose. The potential term becomes
	\begin{align}
		\nonumber V\approx&\int dx\left[|r|A^2+\frac{1}{2}(\partial_xA)^2\right]\\
		\nonumber&+\sum_n\int_{2n}^{2n+1}ds(x_{2n+1}-x_{2n})\left\{
		\frac{2|r|}{(x_{2n+1}-x_{2n})^2}\left(a_{2n+1}-a_{2n}\right)^2
		-\frac{2\sqrt{2}|r|A\left(a_{2n+1}-a_{2n}\right)}{(x_{2n+1}-x_{2n})}
		\right\}\\
		&+\sum_n\int_{2n-1}^{2n}ds(x_{2n}-x_{2n-1})\left\{
		\frac{2|r|}{(x_{2n}-x_{2n-1})^2}\left(b_{2n}-b_{2n-1}\right)^2
		-\frac{2\sqrt{2}|r|A\left(b_{2n}-b_{2n-1}\right)}{(x_{2n}-x_{2n-1})}
		\right\}
	\end{align}
	
	Using Eq.~(\ref{Eq:h_x}), the kinetic energy is expressed by:
	\begin{align}
		\nonumber K=&\frac{1}{6}\sum_n(x_{n+1}-x_n)\left\{\begin{array}{c}
			\left[\partial_{\tau}\theta_{n+1}-(\partial_{\tau}x_{n+1})(-1)^{n+1}m_0-(\partial_{\tau}x_{n+1})h\right]^2\\[1mm]
			+\left[\partial_{\tau}\theta_{n+1}-(\partial_{\tau}x_{n+1})(-1)^{n+1}m_0-(\partial_{\tau}x_{n+1})h\right]\left[\partial_{\tau}\theta_{n}-(\partial_{\tau}x_n)(-1)^{n+1}m_0-(\partial_{\tau}x_n)h\right]\\[1mm]
			+\left[\partial_{\tau}\theta_{n}-(\partial_{\tau}x_n)(-1)^{n+1}m_0-(\partial_{\tau}x_n)h\right]^2
		\end{array}\right\}\\
		\nonumber=&\frac{1}{3}\sum_n\left[(x_{2n+1}-x_{2n})\left(\dot{a}_{2n+1}^2+\dot{a}_{2n}^2+\dot{a}_{2n+1}\dot{a}_{2n}\right)+(x_{2n}-x_{2n-1})\left(\dot{b}_{2n-1}^2+\dot{b}_{2n}^2+\dot{b}_{2n-1}\dot{b}_{2n}\right)\right]+O(a^3,b^3,a^2b,ab^2),
	\end{align}
	where the higher order terms are ignored.

Our goal is to derive an effective action $\mathcal{S}_{A,\text{eff}}[A]$ by integrating out $a_n$ and $b_n$ at the Gaussian level. Formally, we consider
\begin{align}
	\nonumber&\hat{\Phi}^{\dagger}\hat{M}\hat{\Phi}+\hat{\xi}^{\dagger}\hat{\Phi}+\hat{\Phi}^{\dagger}\hat{\xi}\\
	=&\left(\hat{\Phi}^{\dagger}+\hat{\xi}^{\dagger}\hat{M}^{-1}\right)\hat{M}\left(\hat{\Phi}+\hat{M}^{-1}\hat{\xi}\right)-\hat{\xi}^{\dagger}\hat{M}^{-1}\hat{\xi},
\end{align}
where $\hat{\Phi}$ is a $2N$-dimensional vector made of $a_n$ and $b_n$ (with $N$ being the number of domain walls), $\hat{\xi}$ is a $2N$-dimensional vector, and $\hat{M}$ is a $(2N)\times(2N)$ matrix. $\hat{\xi}$ and $\hat{M}$ are functions of $A$. Particularly, $\hat{\xi}$ vanishes when $A\rightarrow 0$. Since we are interested in the correction to $A^2$ order, we drop $O(A^2)$ in $\hat{\xi}$. Similarly, we keep only the $O(A^0)$ terms in $\hat{M}$. Moreover, we drop $\partial_xa_n$ and $\partial_xb_n$ terms because they don't contribute to ac conductivity or stiffness in the limit we are interested in.

In addition, we ignore the fluctuation of the domain size and assume that $x_{n+1}-x_n=\bar{l}$. Such an assumption is valid when the domain size is sufficiently large and the Casimir potential coefficent $\mathcal{C}\gg 1$. The action associated with $K$ is as follows:
	\begin{align}
		\mathcal{S}_K\approx&\frac{\bar{l}}{3}\int d\tau\sum_n\left[\dot{a}_{2n+1}^2+\dot{a}_{2n}^2+\dot{a}_{2n+1}\dot{a}_{2n}+\dot{b}_{2n-1}^2+\dot{b}_{2n}^2+\dot{b}_{2n-1}\dot{b}_{2n}\right]\\
		=&\frac{\bar{l}}{3}\frac{1}{\beta}\sum_{\omega_m}\sum_k\omega_m^2\left[\begin{array}{c}
			\tilde{a}_{1}(-\omega_m,-k)\tilde{a}_{1}(\omega_m,k)+\tilde{a}_{0}(-\omega_m,-k)\tilde{a}_{0}(\omega_m,k)\\[2mm]
			+\frac{1}{2}\tilde{a}_{1}(-\omega_m,-k)\tilde{a}_{0}(\omega_m,k)+\frac{1}{2}\tilde{a}_{0}(-\omega_m,-k)\tilde{a}_{1}(\omega_m,k)\\[2mm]
			+\tilde{b}_{1}(-\omega_m,-k)\tilde{b}_{1}(\omega_m,k)+\tilde{b}_{0}(-\omega_m,-k)\tilde{b}_{0}(\omega_m,k)\\[2mm]
			+\frac{e^{i2k\bar{l}}}{2}\tilde{b}_{1}(-\omega_m,-k)\tilde{b}_{0}(\omega_m,k)+\frac{e^{-i2k\bar{l}}}{2}\tilde{b}_{0}(-\omega_m,-k)\tilde{b}_{1}(\omega_m,k)\\
		\end{array}\right].
	\end{align}
	In the above expressions, we have assumed periodic boundary condition and introduced the Fourier modes as follows:
	\begin{align}
		\tilde{a}_{0}(k)=&\frac{1}{\sqrt{N/2}}\sum_ne^{-i2k\bar{l}n}a_{2n},\,
		\tilde{a}_{1}(k)=\frac{1}{\sqrt{N/2}}\sum_ne^{-i2k\bar{l}n}a_{2n+1},\\
		\tilde{b}_{0}(k)=&\frac{1}{\sqrt{N/2}}\sum_ne^{-i2k\bar{l}n}b_{2n},\,
		\tilde{b}_{1}(k)=\frac{1}{\sqrt{N/2}}\sum_ne^{-i2k\bar{l}n}b_{2n+1},
	\end{align}
	where $N$ is the number of domains. The unit cell size is $2\bar{l}$ and contains 2 $a$'s and 2 $b$'s.
	
	The action associated with $V$ is given by $\mathcal{S}_V'+\mathcal{S}_{A,0}$
	\begin{align}
		\mathcal{S}_V'\!\!\approx\!&\!\int d\tau\sum_n\left\{\begin{array}{c}
			\frac{2|r|}{\bar{l}}\left(a_{2n+1}-a_{2n}\right)^2
			-2\sqrt{2}|r|A\left(a_{2n+1}-a_{2n}\right)\\[2mm]
			+\frac{2|r|}{\bar{l}}\left(b_{2n}-b_{2n-1}\right)^2
			-2\sqrt{2}|r|A\left(b_{2n}-b_{2n-1}\right)
		\end{array}
		\right\}\\[2mm]
		=&\!\frac{1}{\beta}\!\sum_{\omega_m,k}\!\!\left\{\!\!\!\begin{array}{c}
			\frac{2|r|}{\bar{l}}\left[\tilde a_0(-\omega_m,-k)\tilde a_0(\omega_m,k)+\tilde a_1(-\omega_m,-k)\tilde a_1(\omega_m,k)-\tilde a_0(-\omega_m,-k)\tilde a_1(\omega_m,k)-\tilde a_1(-\omega_m,-k)\tilde a_0(\omega_m,k)\right]\\[2mm]
			-\sqrt{2}|r|A(-\omega_n,-k)\left[\tilde a_{1}(\omega_n,k)-\tilde a_{0}(\omega_n,k)\right]-\sqrt{2}|r|\left[\tilde a_{1}(-\omega_n,-k)-\tilde a_{0}(-\omega_n,-k)\right]A(\omega_n,k)\\[2mm]
			+\frac{2|r|}{\bar{l}}\!\left[\tilde b_0(-\omega_m,-k)\tilde b_0(\omega_m,k)\!+\!\tilde b_1(-\omega_m,-k)\tilde b_1(\omega_m,k)\!-\!e^{-i2k\bar{l}}\tilde b_0(-\omega_m,-k)\tilde b_1(\omega_m,k)\!-\!e^{i2k\bar{l}}\tilde b_1(-\omega_m,-k)\tilde b_0(\omega_m,k)\right]\\[2mm]
			-\sqrt{2}|r|A(-\omega_n,-k)\left[e^{-i2k\bar{l}}\tilde b_{1}(\omega_n,k)-\tilde b_{0}(\omega_n,k)\right]-\sqrt{2}|r|\left[e^{i2k\bar{l}}\tilde b_{1}(-\omega_n,-k)-\tilde b_{0}(-\omega_n,-k)\right]A(\omega_n,k)
		\end{array}\!\!\!
		\right\},\\
		\mathcal{S}_{A,0}=&\int d\tau dx\left[|r|A^2+\frac{1}{2}(\partial_xA)^2\right].
	\end{align}
	Our goal is to derive the correction to the $|r|A^2$ term in $\mathcal{S}_{A,0}$.
	
	With the expressions of various actions, we are in the position to integrate out the $a_n$ and $b_n$ variables. The $\mathcal{S}_K+\mathcal{S}_V'+\mathcal{S}_{\text{dis}}$ can be expressed by
	\begin{align}
		\label{Eq:S_K_S_V_S_dis}\mathcal{S}_K+\mathcal{S}_V'+\mathcal{S}_{\text{dis}}
		=\frac{1}{2\beta}\sum_{\omega_m,k}\left[\hat{\Phi}^{\dagger}_{\omega_m,k}\hat{M}(\omega_m,k)\hat{\Phi}_{-\omega_m,-k}+\hat{\xi}^{\dagger}_{\omega_m,k}\hat{\Phi}_{\omega_m,k}+\hat{\Phi}^{\dagger}_{\omega_m,k}\hat{\xi}_{\omega_m,k}\right],
	\end{align}
	where 
	\begin{align}
		\hat{M}(\omega_m,k)=&\frac{2\bar{l}}{3}\omega_m^2\left[\begin{array}{cccc}
			1 & 0 &1/2 & 0 \\
			0 & 1 & 0 & e^{-i2k\bar{l}}/2 \\
			1/2& 0 & 1 & 0 \\
			0 & e^{i2k\bar{l}}/2 & 0 & 1 
		\end{array}
		\right]+\frac{4|r|}{\bar{l}}\left[\begin{array}{cccc}
			1 & 0 &-1 & 0 \\
			0 & 1 & 0 & -e^{-i2k\bar{l}} \\
			-1& 0 & 1 &  0\\
			0 & -e^{i2k\bar{l}} & 0 & 1 
		\end{array}
		\right]+\frac{\gamma}{2m_0}|\omega_m|\left[\begin{array}{cccc}
			1 & -1 &0 & 0 \\
			-1 & 1 & 0 & 0 \\
			0 & 0 & 1 & -1 \\
			0 & 0 & -1 & 1 
		\end{array}
		\right],\\
		\hat{\xi}_{\omega_m,k}=&A(\omega_m,k)\left[\begin{array}{l}
			2\sqrt{2}|r|\\
			2\sqrt{2}|r|\\
			-2\sqrt{2}|r|\\
			-2\sqrt{2}|r|e^{-i2k\bar{l}}
		\end{array}\right],\\
		\hat{\Phi}_{\omega_m,k}=&\left[\begin{array}{r}
			a_0(\omega_m,k)\\ 
			b_0(\omega_m,k)\\
			a_1(\omega_m,k)\\
			b_1(\omega_m,k)
		\end{array}\right].
	\end{align}
	
	\newpage

The effective action for $\tilde{A}$ is given by
\begin{align}
	\nonumber\mathcal{S}_{A,\text{eff}}=&\mathcal{S}_{A,0}-\frac{1}{2\beta}\sum_{\omega_m,k}\hat{\xi}^{\dagger}_{\omega_m,k}\hat{M}^{-1}(\omega_m,k)\hat{\xi}_{\omega_m,k}\\
	\nonumber=&\frac{2\bar{l}|r|}{\beta}\sum_{\omega_m,k}A(-\omega_m,-k)A(\omega_m,k)\\
	&-\frac{1}{2\beta}\sum_{\omega_m,k}\hat{\xi}^{\dagger}_{\omega_m,k}\hat{M}^{-1}(\omega_m,k)\hat{\xi}_{\omega_m,k}.
\end{align}
Note that the unit cell size is $2\bar{l}$.
Using Mathematica and taking the long-wavelength limits, we obtain two asymptotic results
\begin{align}
	&\frac{1}{2}\hat{\xi}^{\dagger}_{\omega_m,k}\hat{M}^{-1}(\omega_m,k)\hat{\xi}_{\omega_m,k}\bigg|_{k= 0,\omega_m\rightarrow 0}
	\rightarrow\frac{16m_0|r|^2\gamma\bar{l}A(-\omega_m,0)A(\omega_m,0)}{8m_0|r|\gamma+\left(8m_0^2|r|+\gamma^2\right)|\omega_m|\bar{l}},\\[2mm]
	&\frac{1}{2}\hat{\xi}^{\dagger}_{\omega_m,k}\hat{M}^{-1}(\omega_m,k)\hat{\xi}_{\omega_m,k}\bigg|_{k= 0,\gamma= 0}
	\rightarrow\frac{48|r|^2\bar{l}}{24|r|+\omega_m^2\bar{l}^2}A(-\omega_m,0)A(\omega_m,0),
\end{align}

Thus, the effective action for $\tilde{A}$ (with $k=0$) is given by
\begin{align}
	\nonumber&\mathcal{S}_{A,\text{eff}}\bigg|_{k=0}
	\equiv\frac{\bar{l}}{\beta}\sum_{\omega_m}Q(\omega_m)\tilde{A}(-\omega_m)\tilde{A}(\omega_m)\\
	=&\frac{2\bar{l}}{\beta}\sum_{\omega_m}|r|\tilde{A}(-\omega_m)\tilde{A}(\omega_m)-\frac{1}{2}\hat{\xi}^{\dagger}_{\omega_m,k}\hat{M}^{-1}(\omega_m,k)\hat{\xi}_{\omega_m,k}\bigg|_{k=0}.
\end{align}
With a nonzero $\gamma$, the low-frequency ac conductivity ($\sigma_{\text{ac}}$) and stiffness ($\rho_s$) are given by
\begin{align}
	\nonumber&\sigma_{\text{ac}}(i\omega_m)\propto\frac{1}{\omega_m}Q(\omega_m)\\
	\rightarrow &\sigma_{\text{ac}}(\omega)\propto\frac{i}{\omega}\left[2|r|-\frac{16m_0|r|^2\gamma}{8m_0|r|\gamma+\left(8m_0^2|r|+\gamma^2\right)(-i\omega)\bar{l}}\right],\\
	\label{Eq:ac_cond}\rightarrow&\text{Re}\left[\sigma_{\text{ac}}(\omega)\right]\propto\frac{16m_0|r|^2\gamma\left(8m_0^2|r|+\gamma^2\right)\bar{l}}{\left(8m_0|r|\gamma\right)^2+\left[\left(8m_0^2|r|+\gamma^2\right)\omega\bar{l}\right]^2},\\
	&\rho_s\propto Q(\omega_m=0)=0.
\end{align}
We note that we have performed analytic continuation from Matsubara frequencies to real frequencies. See \cite{Kamenev2009keldysh} for discussion on the analytic continuation of the phonon drag contribution in the response function. Equation (\ref{Eq:ac_cond}) is consistent with the Drude formula in the low-frequency limit, i.e. the real part of ac conductivity is nonzero for $\omega\rightarrow 0$.

In the absence of phonon drag (i.e., $\gamma=0$), $a_n$ and $b_n$ are decoupled in Eq.~(\ref{Eq:S_K_S_V_S_dis}). As a result, $\hat{M}$ can be block diagonalized. We can easily perform a similar analysis and show that
\begin{align}
	\sigma_{\text{ac}}(\omega)|_{\gamma=0}\propto& \frac{i}{\omega}\left(2|r|-\frac{2|r|^2}{|r|-\omega^2\bar{l}^2/24}\right),\\
	\rho_s|_{\gamma=0}=&0,
\end{align}
resulting in zero superfluid stiffness and the absence of Ohmic response.

\section{Solving Gross-Pitaevski equation}

To verify the linear response theory, we study the problem
using a discretized Gross-Pitaevski equation (GPE), which can simulate bosons in the semiclassical limit.
First, we consider a lattice model given by
\begin{align}
	\nonumber \hat{H}_{\text{Lat}}=&-J_1\sum_j (\psi_{j+1}^* \psi_j+h.c)+J_2\sum_j (\psi_{j+2}^* \psi_j+h.c)+\sum_j (\tilde{U} |\psi_j|^4/2-\mu |\psi_j|^2),
\end{align}
where $J_1>0$ and $J_2>0$ denote the nearest neighbor and the second nearest neighbor hoppings respectively, and $\psi_j$ is the bosonic annihilation operator at site $j$, and $\tilde{U}>0$ is the onsite interaction. The kinetic energy term of the above Hamiltonian leads to a dispersion $\epsilon_k=-2J_1\cos{k}+2J_2\cos{2k}$, which yield two minima for $J_2/J_1>0.25$. We choose $J_2/J_1=0.4$ and $U/J_1=0.8$ for the parameters of our simulations. $\mu$ is chosen so
that $\psi_j=1$ is a ground state of the Hamiltonian. We then solve the Gross-Pitaevskii equations $i (d/dt)\psi_j=\delta \hat{H}_{\text{Lat}}/\delta\psi_j^*$
numerical using the $ode89$ solver in MATLAB from the various initial conditions that we discuss below.

\begin{figure}[t]
	\includegraphics[width=0.3\textwidth]{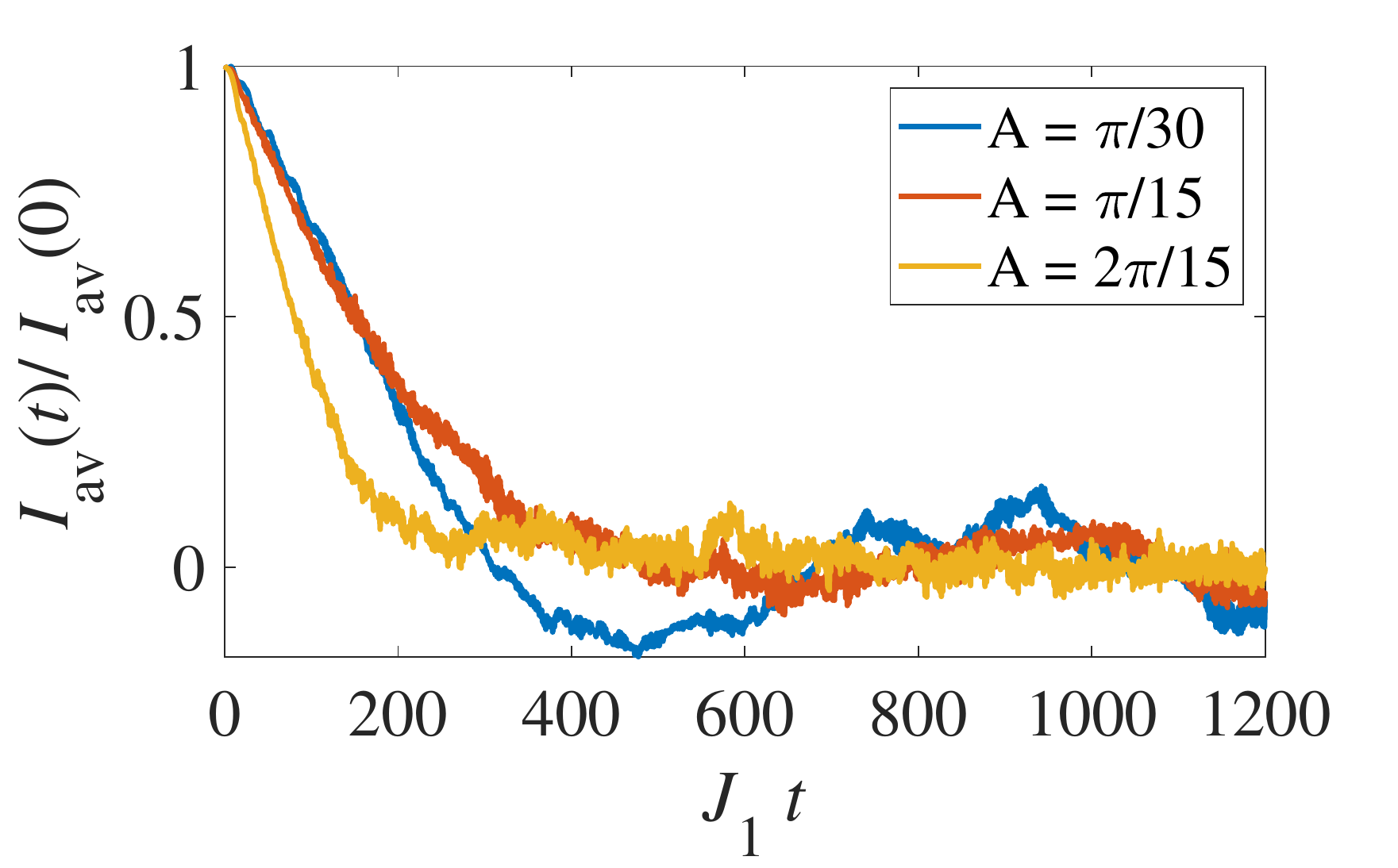}
	\caption{Average current evolution. We prepare initial states with several values of $A$, and then we keep track of the time evolution of average current (normalized to the current at $t=0$) without a vector potential. The results suggest that average current decays to a much smaller value with some oscillation. The relative oscillation amplitude gets smaller for a larger initial value of $A$. $L=1200$ for all the data. The numerical results are obtained by solving Gross-Pitaevskii equations with the $ode89$ solver in MATLAB.}
	\label{Fig:Iav}
\end{figure}

The main goal of our simulation is to confirm the Ohmic response of the finite-temperature states with a few domain-walls.
To do this, we choose initial conditions $\psi_j=e^{i\theta_j}$ together with a choice for the phase-variable $\theta_j$ where the sign of the slope
of $\theta_j$ varies across domain walls in space. In addition, we assume that the system is subject to a large uniform electric field for a short
time, which as discussed in the previous subsection, corresponds to a tilting of the phase profile $\theta_j\rightarrow\theta_j+Aj$.
The ensuing dynamics obtained from the numerical solution of the GPE, shown in Fig.~3(a), confirms the relaxation of the phase profile to a configuration
where the slopes obey the ground state value as time progress through the simulation.

To understand the observable transport consequences of this relaxation we compute the discrete current operator $I_j=-J_1\text{Im}[\psi_{2j+1}^*\psi_{2j}]+J_2\text{Im}[\psi^*_{2j}\psi_{2j-2}+\psi^*_{2j+1}\psi_{2j-1}]$.
This current operator corresponds to a ladder configuration where the sites are paired up into dimers $(2j-1,2j)$ and the current $I_j$ is
between neighboring dimers. In Fig.~3(b), we show the current profiles for a few representative times corresponding to the phase profiles in Fig.~3(b). There are two important messages here. First, the current relaxes, suggesting a non-superfluid behavior. Second, as shown in Fig.~\ref{Fig:Iav}, the average current decreases substantially from the initial value, suggesting a vanishing current in the long-time limit. The decay of current confirms the Ohmic transport as predicted by our linear response theory.

\section{Hydrodynamics near quantum Lifshitz critical point}\label{App:Hydro}

The effective action $S_{\text{eff}}$ in the real-time path integral is given by
\begin{align}
	\nonumber\mathcal{S}=&\!\!\int dtdx\mathcal{L}\\
	\label{Eq:S_LFT}=&\!\!\int dt dx\left[\frac{1}{2}\left(\partial_t\theta\right)^2-\frac{1}{2}\left(\partial_x^2\theta\right)^2-\frac{r}{2}\left(\partial_x\theta\right)^2-u(\partial_x\theta)^4\right],
\end{align}
where $\mathcal{L}$ is the Lagrangian density, and $\theta$ is a real-valued bosonic field. The equation of motion can be derived by the Euler-Lagrange equation (with higher order derivative terms) and is given by
\begin{align}
	\nonumber&\delta \mathcal{S}=0\\
	\rightarrow& 
	\frac{\partial \mathcal{L}}{\partial \theta}-\frac{\partial}{\partial t}\left[\frac{\partial \mathcal{L}}{\partial \left(\partial_t\theta\right)}\right]-\frac{\partial}{\partial x}\left[\frac{\partial \mathcal{L}}{\partial \left(\partial_x\theta\right)}\right]+
	\frac{\partial^2}{\partial x^2}\left[\frac{\partial \mathcal{L}}{\partial \left(\partial_x^2\theta\right)}\right]=0\\
	\nonumber\rightarrow&-\left(\partial_t^2\theta\right)+r\left(\partial_x^2\theta\right)+4u\partial_x\left(\left(\partial_x\theta\right)^3\right)-\left(\partial_x^4\theta\right)=0\\
	\rightarrow&\partial_t\left(\partial_t\theta\right)-\partial_x\left[r\left(\partial_x\theta\right)+4u\left(\partial_x\theta\right)^3-\partial_x^3\theta\right]=0.
\end{align}

The conjugated momentum of $\theta(x)$ is given by
\begin{align}
	\Pi(x)\equiv\frac{\partial \mathcal{L}}{\partial (\partial_t \theta)}=\partial_t\theta.
\end{align}
The Hamiltonian density is given by 
\begin{align}
	\mathcal{H}=\Pi\left(\partial_t\theta\right)-\mathcal{L}=\frac{1}{2}\Pi^2+\frac{1}{2}\left(\partial_x^2\theta\right)^2+\frac{r}{2}\left(\partial_x\theta\right)^2+u(\partial_x\theta)^4.
\end{align}

\subsection{Current and stress tensor}

The physical density and current operators are given by
\begin{align}
	j_t(x,t)=&-\partial_t\theta-A_0,\\
	j_x(x,t)=&r\left(\partial_x\theta\right)+4u\left(\partial_x\theta\right)^3-\partial_x^3\theta,
\end{align}
where $A_0\neq0$ corresponds to a finite density of bosons.
(We define $j_t$ with a minus sign for technical convenience.) The $-A_0$ part of $j_t$ is the zero mode associated with the finite density of the bosons. Thus, $A_0$ does not scale under renormalization. The continuity equation $\partial_t j_t+\partial_xj_x=0$ is ensured by the Euler-Lagrangian equation.
Since the $\theta$-only action is derived in the high density limit, we consider a Lagrangian density with a finite $A_0$ given by
\begin{align}\label{Eq:S_LFT_n}
	\mathcal{L}=\frac{1}{2}\left(\partial_t\theta\right)^2-\frac{1}{2}\left(\partial_x^2\theta\right)^2-\frac{r}{2}\left(\partial_x\theta\right)^2-u(\partial_x\theta)^4+A_0\left(\partial_t\theta\right).
\end{align}
The finite-temperature scaling of sound velocity depends on whether $A_0$ is nonzero. Again, $A_0$ is a temperature-independent parameter because the density should be temperature-independent. Thus, $A_0$ does not scale under renormalization.

To derive the energy-momentum stress tensor, we focus on translation operations. We consider $x^{\mu}\rightarrow x^{\mu}-a^{\mu}$. Alternatively, the field is transformed by 
\begin{align}
	\theta(x)\rightarrow\theta(x+a)=\theta(x)+a^{\mu}\partial_{\mu}\theta(x).
\end{align}
In this case, the Lagrangian density is transformed in the following way, $\mathcal{L}\rightarrow\mathcal{L}+a^{\nu}\partial_{\mu}\left(\delta^{\mu}_{\nu}\mathcal{L}\right)$. We find that
	\begin{align}
		\nonumber&a^0\left\{
		\partial_t\left[\frac{\partial\mathcal{L}}{\partial \left(\partial_t\theta\right)}\left(\partial_t\theta\right)-\mathcal{L}\right]
		+\partial_x\left[\frac{\partial\mathcal{L}}{\partial \left(\partial_x\theta\right)}\left(\partial_t\theta\right)+\frac{\partial\mathcal{L}}{\partial \left(\partial_x^2\theta\right)}\left(\partial_x\partial_t\theta\right)-\partial_x\left(\frac{\partial\mathcal{L}}{\partial \left(\partial_x^2\theta\right)}\right)\left(\partial_t\theta\right)\right]	
		\right\}\\
		+&a^1\left\{
		\partial_t\left[\frac{\partial\mathcal{L}}{\partial \left(\partial_t\theta\right)}\left(\partial_x\theta\right)\right]
		+\partial_x\left[\frac{\partial\mathcal{L}}{\partial \left(\partial_x\theta\right)}\left(\partial_x\theta\right)+\frac{\partial\mathcal{L}}{\partial \left(\partial_x^2\theta\right)}\left(\partial_x^2\theta\right)-\partial_x\left(\frac{\partial\mathcal{L}}{\partial \left(\partial_x^2\theta\right)}\right)\left(\partial_x\theta\right)-\mathcal{L}\right]	
		\right\}=0
	\end{align}

	The stress tensor is defined by
	\begin{align}
		T_{tt}=&\frac{\partial\mathcal{L}}{\partial \left(\partial_t\theta\right)}\left(\partial_t\theta\right)-\mathcal{L}=\frac{1}{2}\left(\partial_t\theta\right)^2+\frac{1}{2}\left(\partial_x^2\theta\right)^2+\frac{r}{2}\left(\partial_x\theta\right)^2+u(\partial_x\theta)^4=\mathcal{H},\\
		T_{xt}=&\frac{\partial\mathcal{L}}{\partial \left(\partial_t\theta\right)}\left(\partial_x\theta\right)=\left(\partial_t\theta\right)\left(\partial_x\theta\right)+A_0\left(\partial_x\theta\right),\\
		T_{tx}=&\frac{\partial\mathcal{L}}{\partial \left(\partial_x\theta\right)}\left(\partial_t\theta\right)
		+\frac{\partial\mathcal{L}}{\partial \left(\partial_x^2\theta\right)}\left(\partial_x\partial_t\theta\right)-\partial_x\left(\frac{\partial\mathcal{L}}{\partial \left(\partial_x^2\theta\right)}\right)\left(\partial_t\theta\right)\\
		=&\left[-r\left(\partial_x\theta\right)-4u\left(\partial_x\theta\right)^3+\left(\partial_x^3\theta\right)\right]\left(\partial_t\theta\right)-\left(\partial_x^2\theta\right)\left(\partial_x\partial_t\theta\right),\\
		T_{xx}=&\frac{\partial\mathcal{L}}{\partial \left(\partial_x\theta\right)}\left(\partial_x\theta\right)
		+\frac{\partial\mathcal{L}}{\partial \left(\partial_x^2\theta\right)}\left(\partial_x^2\theta\right)-\partial_x\left(\frac{\partial\mathcal{L}}{\partial \left(\partial_x^2\theta\right)}\right)\left(\partial_x\theta\right)-\mathcal{L}\\
		=&-\frac{1}{2}\left(\partial_t\theta\right)^2-\frac{1}{2}\left(\partial_x^2\theta\right)^2-\frac{r}{2}\left(\partial_x\theta\right)^2-3u(\partial_x\theta)^4+\left(\partial_x^3\theta\right)\left(\partial_x\theta\right)-A_0\left(\partial_t\theta\right),
	\end{align}
	We can easily check that
	$\partial_tT_{tt}+\partial_xT_{tx}=0$ and $\partial_tT_{xt}+\partial_xT_{xx}=0$.

\subsection{Gaussian fixed point}

Now, we determine the scaling dimensions of operators and RG eigenvalues of coupling constants in the action given by Eq.~(\ref{Eq:S_LFT}). First, we focus on the scaling near the Gaussian fixed point, corresponding to
\begin{align}
	[x]=-1,\,\,
	[t]=-z,\,\,
	[\theta]=-\frac{\epsilon}{2},\,\,
	y_r=2,\,\,\,
	y_u=\epsilon,
\end{align}
where $\epsilon=2-d=1$ in our case. Notice that $[\theta]$ does not explicitly depend on $y_r$ or $y_u$. Then, we examine the scaling dimensions of $T_{\mu\nu}$ and $j_{\mu}$ in the following.

In our theory with a finite density, $A_0$ is a constant that does not scale under RG. Thus, $[A_0]=0$. We focus on the leading scaling dimension of $[T_{xt}]$ and $[T_{xx}]$, which are given by $[\partial_x\theta]$ and $[\partial_t\theta]$ respectively. 
Based on dimensional analysis, we derive
\begin{align}
	\nonumber	y_{\beta}=&-2,\,\,
	[T_{tt}]=3,\,\,
	[T_{xt}]=\frac{1}{2},\,\,[T_{tx}]=4,\\
	\label{Eq:Scaling}[T_{xx}]=&\frac{3}{2},\,\,
	[j_t]=0,\,\,
	[j_x]=\frac{5}{2}.
\end{align}
Note that $[T_{tt}]=z+d$ holds for quantum Lifshitz field theory \cite{Hoyos2013lifshitz,Hoyos2014lifshitz}. Here, $z=2$ and $d=1$.\\

\subsection{Interacting fixed point}

We are interested in the results near the interacting fixed point. We first recall the RG equations \cite{Yang2004} at order $O(\epsilon)$ given by
\begin{align}
	\frac{d r}{dl}=&2r+\frac{3u}{\pi}\left(\Lambda^2-\frac{1}{2}r\right),\\
	\frac{d u}{dl}=&\epsilon u-\frac{9}{2\pi}u^2.
\end{align}
The fixed point is given by $(r^*,u^*)=(-\epsilon\Lambda^2/3,2\pi\epsilon/9)$ (upto linear-in-$\epsilon$ order). The RG dimensions for $r$ and $w=\frac{9\Lambda^2}{4\pi(\epsilon-3)}r+u$ are $y_r=2-\epsilon/3$ and $y_w=\epsilon-2$. Since $u$ is no longer the eigen direction of the RG flow, the scaling of $u$ has two contributions. At the end of the calculations, we will set $\epsilon=1$.

Now we discuss the scaling dimensions of $T_{\mu\nu}$ and $j_{\mu}$ near the interacting fixed point. We can easily show that $j_t$, $[T_{tt}]$, $[T_{xt}]$ are unchanged. Then, we use conservation laws and obtain the same scaling results for $j_x$, $[T_{tx}]$, $[T_{xx}]$.
Therefore, the scaling dimensions of $T_{\mu\nu}$ and $j_{\mu}$ remain the same as that in the Gaussian fixed point.

\subsection{Free energy in quantum critical regime}

To construct a hydrodynamic theory, we consider a free energy $F[\mu,\beta,v]$ defined by
\begin{align}
	e^{-\beta F[s,\mu,\beta,v]}=e^{-Lf[s,\mu,\beta,v]}=\tr\left[e^{-\beta\int dx\left(sT_{tt}-vT_{xt}-\mu j_t\right)}\right],
\end{align}
where $\beta$ is the inverse temperature, $s$ is a dimensionless parameter ($s$ is set to 1 at the end of calculations), $v$ is the center of mass velocity, $\mu$ is the chemical potential, $L$ is the system size, and $f$ is the reduced free energy density. 
In this formulation, $H=\int dx T_{tt}$ (energy), $P=\int dx T_{xt}$ (momentum), and $N=\int dx j_t$ (particle number) are the conserved quantity of the theory. Our goal here is to extract the scaling behavior of various hydrodynamic quantities in the quantum critical scaling region defined by $\xi_T<\xi$ ($\xi_T$ is the thermal wavelength and $\xi$ is the correlation length of the theory).\\

We note that the dipole moment conservation is not explicitly implemented in this framework. The reason is that we focus on the interacting fixed point, which does not contain exact dipole moment conservation. Using this present framework, the Guassian fixed point and interacting fixed point yield exact the same scaling behavior. However, the true hydrodynamics of Lifshitz point should incorporate the dipole moment conservation in constructing the free energy functional. We do not explore the hydrodynamics of Lifshitz point incorporating the dipole conservation in this work.

In the quantum critical regime, the temperature dependence can be determined by the hyper-scaling. The
reduced free energy density near the critical point obeys the following scaling relation
\begin{align}
	f[s,\mu,\beta,v]=b^{-d}f[s,\mu b^{y_{\mu}},\beta b^{-z},v b^{y_v}],
\end{align}
where $b$ is the scale parameter, $z=2$ is the dynamic exponent, $y_v$ is the scaling eigenvalue of $v$, and $y_{\mu}$ is the scaling eigenvalue of $\mu$. We can fix $\beta b^{-z}$ to a constant and rewrite the expression in terms of $T=1/\beta$ as follows
\begin{align}
	f[s,\mu,\beta,v]=T^{\frac{d}{z}}\Phi\left[s,\frac{\mu}{T^{\frac{y_{\mu}}{z}}},\frac{v}{T^{\frac{y_{v}}{z}}}\right],
\end{align}
where $\Phi$ is a universal scaling function.
Recall that $\beta F=Lf$, we obtain the free energy density
\begin{align}
	F[s,\mu,1/T,v]/L=T^{1+\frac{d}{z}}\Phi\left[s,\frac{\mu}{T^{\frac{y_{\mu}}{z}}},\frac{v}{T^{\frac{y_{v}}{z}}}\right].
\end{align}

With the expression of the free energy density, we can derive the following expectation values:  
\begin{align}
	\langle T_{tt}\rangle=&\frac{1}{L}\left(\frac{\partial F}{\partial s}\right)_{v,\mu,T}\bigg|_{s=1}\sim T^{1+\frac{d}{z}}=T^{3/2}\\
	\langle T_{xt}\rangle=&-\frac{1}{L}\left(\frac{\partial F}{\partial v}\right)_{T,\mu,s}\bigg|_{s=1}\sim T^{1+\frac{d-y_v}{z}}=T^{1/4},\\
	\langle j_{t}\rangle=&-\frac{1}{L}\left(\frac{\partial F}{\partial \mu}\right)_{T,v,s}\bigg|_{s=1}\sim T^{1+\frac{d-y_{\mu}}{z}}=T^{0},
\end{align}
where we have used $y_v=5/2$ and $y_{\mu}=3$. We also assumed the translation symmetry and analyticity in the scaling function. Using the scaling results in Eq.~(\ref{Eq:Scaling}), we derive that
\begin{align}
	\langle T_{tx}\rangle\sim T^2,\,\,\langle T_{xx}\rangle\sim T^{3/4},\,\,\langle j_x\rangle\sim T^{1/4}.
\end{align}
To derive $[j_x]$, we use $[\delta j_t]\sim T^{3/4}$ as the $A_0$ term does not contribute.

Now, we are in the position to determine the scaling of various thermodynamic quantities. These thermodynamic quantities are given by
\begin{align}
	\delta \langle T_{tt}\rangle/\delta\beta=&C_{\beta}\sim T^{5/2},\\
	\delta \langle T_{tt}\rangle/\delta\mu=&C_{\mu}\sim T^{0},\\
	\delta \langle T_{xx}\rangle/\delta\beta=&K_{\beta}\sim T^{7/4},\\
	\delta \langle T_{xx}\rangle/\delta\mu=&K_{\mu}\sim T^{-3/4},\\
	\delta \langle j_t\rangle/\delta\beta=&\chi_{\beta}\sim T^{7/4},\\
	\delta \langle j_t\rangle/\delta\mu=&\chi_{\mu}\sim T^{-3/4},\\
	\delta \langle T_{xt}\rangle/\delta v=&\rho_M\sim T^{-1},\\
	\delta \langle T_{tx}\rangle/\delta v=&\rho_E\sim T^{3/4},\\
	\delta \langle j_x\rangle/\delta v=&\rho_Q\sim T^{0},
\end{align}
where we have used $[\delta j_t]\sim T^{3/4}$ as the $A_0$ term does not contribute.
Again, we have used the scaling function $\Phi$ and the conservation laws for energy, momentum, and charge.

Note that $C_{\beta}\neq C_T=\delta \langle T_{tt}\rangle/\delta T\sim T^{1/2}$ ($C_T\sim T^{1/2}$ is consistent with Ref.~\cite{Yang2004}). $\rho_M\sim1/T$ is also consistent with the magnetic susceptibility in Ref.~\cite{Yang2004}. The superfluid stiffness $\rho_s=\delta\langle j_x\rangle/\delta A$, where $A$ is the vector potential. Since $\delta A\sim \delta(\partial_x\theta)$ (as $\delta A$ is associated with a twist in the boundary condition), $\delta A\sim A_0^{-1}\delta \langle T_{xt}\rangle$. Thus, the superfluid stiffness can be expressed by $\rho_s=A_0\frac{\delta\langle j_x\rangle}{\delta \langle T_{xt}\rangle}=A_0\frac{\rho_Q}{\rho_M}\sim T$.

%For interacting bosons, $\rho_M$ is the inertia, which is inversely proportional to the superfluid stiffness. Thus, we obtain that $\rho_s\sim\rho_M^{-1}\sim T$, which vanishes at zero temperature.

\subsection{Sound velocity}

The conservation laws for energy, momentum, and charge are given by
\begin{align}
	\partial_t\langle T_{tt}\rangle+\partial_x\langle T_{tx}\rangle=&0,\\
	\partial_t\langle T_{xt}\rangle+\partial_x\langle T_{xx}\rangle=&0,\\
	\partial_t\langle j_t\rangle+\partial_x\langle j_x\rangle=&0.
\end{align}
Using the thermodynamic relations, we obtain
\begin{subequations}\label{Eq:TJ_to_C_rho_chi}
	\begin{align}
		\delta\langle T_{tt}\rangle=&C_{\beta}\delta\beta+C_{\mu}\delta\mu,\\
		\delta\langle T_{tx}\rangle=&\rho_E\delta v,\\
		\delta\langle T_{xt}\rangle=&\rho_M\delta v,\\
		\delta\langle T_{xx}\rangle=&K_{\beta}\delta\beta+K_{\mu}\delta\mu,\\
		\delta\langle j_t\rangle=&\chi_{\mu}\delta\mu+\chi_{\beta}\delta \beta,\\
		\delta\langle j_x\rangle=&\rho_Q\delta v.
	\end{align}
\end{subequations}
With Eq.~(\ref{Eq:TJ_to_C_rho_chi}), the conservation laws can be recast as follows:
\begin{align}
	\left[\begin{array}{ccc}
		C_{\mu} & 0 & C_{\beta}\\
		0 & \rho_M & 0\\
		\chi_{\mu} & 0& \chi_{\beta}
	\end{array}
	\right]\partial_t\left[\begin{array}{c}
		\mu\\
		v\\
		\beta
	\end{array}\right]
	+
	\left[\begin{array}{ccc}
		0 & \rho_E & 0\\
		K_{\mu} & 0 & K_{\beta}\\
		0 & \rho_Q& 0
	\end{array}
	\right]\partial_x\left[\begin{array}{c}
		\mu\\
		v\\
		\beta
	\end{array}\right]=0
\end{align}

Now, we use the property
\begin{align}
	\nonumber&\delta\left[\begin{array}{c}
		\langle T_{tt}\rangle	\\
		\langle T_{xt}\rangle	\\
		\langle j_t\rangle	
	\end{array}
	\right]=\left[\begin{array}{ccc}
		C_{\mu} & 0 & C_{\beta}\\
		0 & \rho_M & 0\\
		\chi_{\mu} & 0& \chi_{\beta}
	\end{array}
	\right]\delta\left[\begin{array}{c}
		\mu\\
		v\\
		\beta
	\end{array}\right]\\
	\rightarrow&
	\delta\left[\begin{array}{c}
		\mu\\
		v\\
		\beta
	\end{array}\right]=	
	\left[\begin{array}{ccc}
		C_{\mu} & 0 & C_{\beta}\\
		0 & \rho_M & 0\\
		\chi_{\mu} & 0& \chi_{\beta}
	\end{array}
	\right]^{-1}\delta\left[\begin{array}{c}
		\langle T_{tt}\rangle	\\
		\langle T_{xt}\rangle	\\
		\langle j_t\rangle	
	\end{array}
	\right].
\end{align}

We find that the conservation law can be expressed by
\begin{align}\label{Eq:wave}
	\partial_t\left[\begin{array}{c}
		\epsilon\\
		p\\
		n
	\end{array}\right]=\left[\begin{array}{ccc}
		0 & M_4 & 0\\
		M_3 & 0 & M_2\\
		0 & M_1 & 0
	\end{array}\right]
	\partial_x\left[\begin{array}{c}
		\epsilon\\
		p\\
		n
	\end{array}\right],
\end{align}
where $\epsilon=\langle T_{tt}\rangle$ is the energy density, $p=\langle T_{xt}\rangle$ is the momentum density, $n=\langle j_t\rangle$ is the number density, and $(M_1,M_2,M_3,M_4)=\left(\rho_Q/\rho_M,\frac{C_{\mu}K_{\beta}-C_{\beta}K_{\mu}}{C_{\mu}\chi_{\beta}-C_{\beta}\chi_{\mu}},\frac{K_{\mu}\chi_{\beta}-K_{\beta}\chi_{\mu}}{C_{\mu}\chi_{\beta}-C_{\beta}\chi_{\mu}},\rho_E/\rho_M\right)$. The eigenvalues of the matrix in Eq.~(\ref{Eq:wave}) correspond to the sound velocities. We find that $v=0,\pm v_s$, where
\begin{align}
	v_s=\sqrt{M_1M_2+M_3M_4}.
\end{align}
The corresponding normal modes are given by
\begin{align}
	\hat{n}_0\propto\left[\begin{array}{c}
		-M_2\\
		0\\
		M_3
	\end{array}
	\right],\,\,\hat{n}_{\pm}\propto
	\left[\begin{array}{c}
		M_4\\
		\pm v_s\\
		M_1
	\end{array}
	\right].
\end{align}

Based on the scaling analysis above, $M_1\sim T$, $M_2\sim T^{0}$, $M_3\sim T^{-3/4}$, and $M_4\sim T^{7/4}$. Thus, $M_1M_2\sim T$ and $M_3M_4\sim T$. Thus, the sound velocity $v_s\sim T^{1/2}$.

\subsection{Digression: Luttinger liquid case}

Here, we briefly discuss the conventional interacting bosons with $k^2$ dispersion. This case corresponds to a large positive $r$, which corresponds to the Luttinger liquid ($z=1$). The action is given by
\begin{align}\label{Eq:S_LL}
	\mathcal{L}'=\frac{1}{2}\left(\partial_t\theta\right)^2-\frac{r}{2}\left(\partial_x\theta\right)^2+A_0\left(\partial_t\theta\right).
\end{align}
The scaling dimensions and the scaling eigenvalues of couplings are as follows: $[\theta]=0$, $y_r=0$, $y_{A_0}=1$, $[j_t]=0$, $[j_x]=1$, $[T_{tt}]=2$, $[T_{xt}]=1$. We can easily show a temperature-independent superfluid stiffness $\rho_s\sim T^0$, which is consistent with Luttinger liquid situation.

\subsection{Digression: Classical gases with Lifshitz dispersion}

It is also interesting to know the finite-temperature behavior in the classical gases with Lifshitz dispersion. To study this, we construct the partition function of $N$ classical gases with Lifshitz dispersion as follows:
\begin{align}
	Z=&\frac{1}{N!}\int\prod_{i=1}^N\left[\frac{dx_idk_i}{2\pi}e^{-\beta\left(k_i^4-vk_i\right)}\right]
	=\frac{1}{N!}\left[L\beta^{-1/4}\Phi\left(v\beta^{3/4}\right)\right]^N,
\end{align}
where $\beta=1/T$ is the inverse temperature and $v$ is the velocity. We will set the $v$ term to zero at the end of calculations.

The free energy is given by $F=-T\ln Z\approx-TN\ln\left[\frac{Le}{N}T^{1/4}\Phi\left(v/T^{3/4}\right)\right]$. The finite-temperature scaling of inertia can be obtained by $\frac{\partial^2 F}{\partial v^2}\sim T^{-1/2}$. Note that the result is qualitatively similar to the interacting quantum Lifshitz theory, but the exponent is different.

In addition, the sound velocity scaling can be derived. To do this, we obtaine the expression of entropy through $S=-\frac{\partial F}{\partial T}$ and set $v=0$. The entropy can be express as a function of $LT^{1/4}/N$. Then, we can easily derive the isentropic bulk modulus $K_s\sim T$, and the sound velocity $v_s\sim T^{1/2}$. Note that the sound velocity gives the same finite-temperature scaling as the interacting quantum Lifshitz theory despite the differences in detail.

The classical theory here is an approximate description for the interacting Lifshitz theory at sufficiently high temperature. Our results show that the inertia finite-temperature scaling is different in the high-temperature ($\rho_M\sim T^{-1/2}$) and in the quantum critical regime ($\rho_M\sim T^{-1}$), while the sound velocity has the same finite-temperature scaling ($v_s\sim T^{1/2}$) in both regimes.

\section{$\mu/\epsilon_0\ll1$ limit}

The 1D bosons with a double-well dispersion realize a rich phase diagram with two tunable dimensionless parameters, $\gamma=U/\left(B\rho_0\right)$ (with $\rho_0$ being the density) and  $\delta=\mu/\epsilon_0$. $\gamma$ is the dimensionless interaction parameter analogous to the dimensionless parameter in the Lieb-Liniger model \cite{LiebLiniger}. Different from the Lieb-Liniger model, a quantum phase transition separates $\gamma\ll 1$ and $\gamma\gg 1$ regimes because of the $Z_2$ character of the double-well dispersion.

In this work, we focus only on $\delta\gg1$ and ignore the fluctuation in density. We anticipate similar constrained domain-wall motion in the general cases, including $\delta\ll 1$ (requiring both density and phase fields). The main difference is that the domain walls in the $\delta\ll 1$ regime feature inevitable density fluctuations (related to the phase staircases) \cite{Liu2016}. The low-temperature symmetry-broken states still have the emergent dipole moment conservation, and much of our analysis based on long-wavelength theory still applies, e.g., long-wavelength phonons are confined in each domain and phonon drag. Therefore, we argue that the superfluid with double-well dispersion generically manifests robust domain walls and slow dynamics in the symmetry-broken regime \cite{Cole2019emergent}.


\begin{thebibliography}{81}
	\expandafter\ifx\csname natexlab\endcsname\relax\def\natexlab#1{#1}\fi
	\expandafter\ifx\csname bibnamefont\endcsname\relax
	\def\bibnamefont#1{#1}\fi
	\expandafter\ifx\csname bibfnamefont\endcsname\relax
	\def\bibfnamefont#1{#1}\fi
	\expandafter\ifx\csname citenamefont\endcsname\relax
	\def\citenamefont#1{#1}\fi
	\expandafter\ifx\csname url\endcsname\relax
	\def\url#1{\texttt{#1}}\fi
	\expandafter\ifx\csname urlprefix\endcsname\relax\def\urlprefix{URL }\fi
	\providecommand{\bibinfo}[2]{#2}
	\providecommand{\eprint}[2][]{\url{#2}}
	
	\bibitem[{\citenamefont{Stenger et~al.}(1998)\citenamefont{Stenger, Inouye,
			Stamper-Kurn, Miesner, Chikkatur, and Ketterle}}]{Stenger1998spin}
	\bibinfo{author}{\bibfnamefont{J.}~\bibnamefont{Stenger}},
	\bibinfo{author}{\bibfnamefont{S.}~\bibnamefont{Inouye}},
	\bibinfo{author}{\bibfnamefont{D.}~\bibnamefont{Stamper-Kurn}},
	\bibinfo{author}{\bibfnamefont{H.-J.} \bibnamefont{Miesner}},
	\bibinfo{author}{\bibfnamefont{A.}~\bibnamefont{Chikkatur}},
	\bibnamefont{and} \bibinfo{author}{\bibfnamefont{W.}~\bibnamefont{Ketterle}},
	\bibinfo{journal}{Nature} \textbf{\bibinfo{volume}{396}},
	\bibinfo{pages}{345} (\bibinfo{year}{1998}).
	
	\bibitem[{\citenamefont{Greiner et~al.}(2002)\citenamefont{Greiner, Mandel,
			Esslinger, H{\"a}nsch, and Bloch}}]{Greiner2002quantum}
	\bibinfo{author}{\bibfnamefont{M.}~\bibnamefont{Greiner}},
	\bibinfo{author}{\bibfnamefont{O.}~\bibnamefont{Mandel}},
	\bibinfo{author}{\bibfnamefont{T.}~\bibnamefont{Esslinger}},
	\bibinfo{author}{\bibfnamefont{T.~W.} \bibnamefont{H{\"a}nsch}},
	\bibnamefont{and} \bibinfo{author}{\bibfnamefont{I.}~\bibnamefont{Bloch}},
	\bibinfo{journal}{Nature} \textbf{\bibinfo{volume}{415}}, \bibinfo{pages}{39}
	(\bibinfo{year}{2002}).
	
	\bibitem[{\citenamefont{Kinoshita et~al.}(2004)\citenamefont{Kinoshita, Wenger,
			and Weiss}}]{Kinoshita2004}
	\bibinfo{author}{\bibfnamefont{T.}~\bibnamefont{Kinoshita}},
	\bibinfo{author}{\bibfnamefont{T.}~\bibnamefont{Wenger}}, \bibnamefont{and}
	\bibinfo{author}{\bibfnamefont{D.~S.} \bibnamefont{Weiss}},
	\bibinfo{journal}{Science} \textbf{\bibinfo{volume}{305}},
	\bibinfo{pages}{1125} (\bibinfo{year}{2004}).
	
	\bibitem[{\citenamefont{Kinoshita et~al.}(2006)\citenamefont{Kinoshita, Wenger,
			and Weiss}}]{Kinoshita2006}
	\bibinfo{author}{\bibfnamefont{T.}~\bibnamefont{Kinoshita}},
	\bibinfo{author}{\bibfnamefont{T.}~\bibnamefont{Wenger}}, \bibnamefont{and}
	\bibinfo{author}{\bibfnamefont{D.~S.} \bibnamefont{Weiss}},
	\bibinfo{journal}{Nature} \textbf{\bibinfo{volume}{440}},
	\bibinfo{pages}{900} (\bibinfo{year}{2006}).
	
	\bibitem[{\citenamefont{Sadler et~al.}(2006)\citenamefont{Sadler, Higbie,
			Leslie, Vengalattore, and Stamper-Kurn}}]{Sadler2006spontaneous}
	\bibinfo{author}{\bibfnamefont{L.}~\bibnamefont{Sadler}},
	\bibinfo{author}{\bibfnamefont{J.}~\bibnamefont{Higbie}},
	\bibinfo{author}{\bibfnamefont{S.}~\bibnamefont{Leslie}},
	\bibinfo{author}{\bibfnamefont{M.}~\bibnamefont{Vengalattore}},
	\bibnamefont{and}
	\bibinfo{author}{\bibfnamefont{D.}~\bibnamefont{Stamper-Kurn}},
	\bibinfo{journal}{Nature} \textbf{\bibinfo{volume}{443}},
	\bibinfo{pages}{312} (\bibinfo{year}{2006}).
	
	\bibitem[{\citenamefont{Lin et~al.}(2011)\citenamefont{Lin,
			Jim{\'e}nez-Garc{\'\i}a, and Spielman}}]{Lin2011spin}
	\bibinfo{author}{\bibfnamefont{Y.-J.} \bibnamefont{Lin}},
	\bibinfo{author}{\bibfnamefont{K.}~\bibnamefont{Jim{\'e}nez-Garc{\'\i}a}},
	\bibnamefont{and} \bibinfo{author}{\bibfnamefont{I.~B.}
		\bibnamefont{Spielman}}, \bibinfo{journal}{Nature}
	\textbf{\bibinfo{volume}{471}}, \bibinfo{pages}{83} (\bibinfo{year}{2011}).
	
	\bibitem[{\citenamefont{Zhang et~al.}(2012)\citenamefont{Zhang, Ji, Chen,
			Zhang, Du, Yan, Pan, Zhao, Deng, Zhai et~al.}}]{Zhang2012}
	\bibinfo{author}{\bibfnamefont{J.-Y.} \bibnamefont{Zhang}},
	\bibinfo{author}{\bibfnamefont{S.-C.} \bibnamefont{Ji}},
	\bibinfo{author}{\bibfnamefont{Z.}~\bibnamefont{Chen}},
	\bibinfo{author}{\bibfnamefont{L.}~\bibnamefont{Zhang}},
	\bibinfo{author}{\bibfnamefont{Z.-D.} \bibnamefont{Du}},
	\bibinfo{author}{\bibfnamefont{B.}~\bibnamefont{Yan}},
	\bibinfo{author}{\bibfnamefont{G.-S.} \bibnamefont{Pan}},
	\bibinfo{author}{\bibfnamefont{B.}~\bibnamefont{Zhao}},
	\bibinfo{author}{\bibfnamefont{Y.-J.} \bibnamefont{Deng}},
	\bibinfo{author}{\bibfnamefont{H.}~\bibnamefont{Zhai}}, \bibnamefont{et~al.},
	\bibinfo{journal}{Phys. Rev. Lett.} \textbf{\bibinfo{volume}{109}},
	\bibinfo{pages}{115301} (\bibinfo{year}{2012}),
	\urlprefix\url{https://link.aps.org/doi/10.1103/PhysRevLett.109.115301}.
	
	\bibitem[{\citenamefont{Nguyen et~al.}(2014)\citenamefont{Nguyen, Dyke, Luo,
			Malomed, and Hulet}}]{Nguyen2014collisions}
	\bibinfo{author}{\bibfnamefont{J.~H.} \bibnamefont{Nguyen}},
	\bibinfo{author}{\bibfnamefont{P.}~\bibnamefont{Dyke}},
	\bibinfo{author}{\bibfnamefont{D.}~\bibnamefont{Luo}},
	\bibinfo{author}{\bibfnamefont{B.~A.} \bibnamefont{Malomed}},
	\bibnamefont{and} \bibinfo{author}{\bibfnamefont{R.~G.} \bibnamefont{Hulet}},
	\bibinfo{journal}{Nature Physics} \textbf{\bibinfo{volume}{10}},
	\bibinfo{pages}{918} (\bibinfo{year}{2014}).
	
	\bibitem[{\citenamefont{Beeler et~al.}(2013)\citenamefont{Beeler, Williams,
			Jimenez-Garcia, LeBlanc, Perry, and Spielman}}]{Beeler2013spin}
	\bibinfo{author}{\bibfnamefont{M.~C.} \bibnamefont{Beeler}},
	\bibinfo{author}{\bibfnamefont{R.~A.} \bibnamefont{Williams}},
	\bibinfo{author}{\bibfnamefont{K.}~\bibnamefont{Jimenez-Garcia}},
	\bibinfo{author}{\bibfnamefont{L.~J.} \bibnamefont{LeBlanc}},
	\bibinfo{author}{\bibfnamefont{A.~R.} \bibnamefont{Perry}}, \bibnamefont{and}
	\bibinfo{author}{\bibfnamefont{I.~B.} \bibnamefont{Spielman}},
	\bibinfo{journal}{Nature} \textbf{\bibinfo{volume}{498}},
	\bibinfo{pages}{201} (\bibinfo{year}{2013}).
	
	\bibitem[{\citenamefont{Parker et~al.}(2013)\citenamefont{Parker, Ha, and
			Chin}}]{Parker2013direct}
	\bibinfo{author}{\bibfnamefont{C.~V.} \bibnamefont{Parker}},
	\bibinfo{author}{\bibfnamefont{L.-C.} \bibnamefont{Ha}}, \bibnamefont{and}
	\bibinfo{author}{\bibfnamefont{C.}~\bibnamefont{Chin}},
	\bibinfo{journal}{Nature Physics} \textbf{\bibinfo{volume}{9}},
	\bibinfo{pages}{769} (\bibinfo{year}{2013}).
	
	\bibitem[{\citenamefont{Jim\'enez-Garc\'{\i}a
			et~al.}(2015)\citenamefont{Jim\'enez-Garc\'{\i}a, LeBlanc, Williams, Beeler,
			Qu, Gong, Zhang, and Spielman}}]{JimenezGarcia2015}
	\bibinfo{author}{\bibfnamefont{K.}~\bibnamefont{Jim\'enez-Garc\'{\i}a}},
	\bibinfo{author}{\bibfnamefont{L.~J.} \bibnamefont{LeBlanc}},
	\bibinfo{author}{\bibfnamefont{R.~A.} \bibnamefont{Williams}},
	\bibinfo{author}{\bibfnamefont{M.~C.} \bibnamefont{Beeler}},
	\bibinfo{author}{\bibfnamefont{C.}~\bibnamefont{Qu}},
	\bibinfo{author}{\bibfnamefont{M.}~\bibnamefont{Gong}},
	\bibinfo{author}{\bibfnamefont{C.}~\bibnamefont{Zhang}}, \bibnamefont{and}
	\bibinfo{author}{\bibfnamefont{I.~B.} \bibnamefont{Spielman}},
	\bibinfo{journal}{Phys. Rev. Lett.} \textbf{\bibinfo{volume}{114}},
	\bibinfo{pages}{125301} (\bibinfo{year}{2015}),
	\urlprefix\url{https://link.aps.org/doi/10.1103/PhysRevLett.114.125301}.
	
	\bibitem[{\citenamefont{Clark et~al.}(2016)\citenamefont{Clark, Feng, and
			Chin}}]{Clark2016}
	\bibinfo{author}{\bibfnamefont{L.~W.} \bibnamefont{Clark}},
	\bibinfo{author}{\bibfnamefont{L.}~\bibnamefont{Feng}}, \bibnamefont{and}
	\bibinfo{author}{\bibfnamefont{C.}~\bibnamefont{Chin}},
	\bibinfo{journal}{Science} \textbf{\bibinfo{volume}{354}},
	\bibinfo{pages}{606} (\bibinfo{year}{2016}).
	
	\bibitem[{\citenamefont{Putra et~al.}(2020)\citenamefont{Putra,
			Salces-C\'arcoba, Yue, Sugawa, and Spielman}}]{Putra2020}
	\bibinfo{author}{\bibfnamefont{A.}~\bibnamefont{Putra}},
	\bibinfo{author}{\bibfnamefont{F.}~\bibnamefont{Salces-C\'arcoba}},
	\bibinfo{author}{\bibfnamefont{Y.}~\bibnamefont{Yue}},
	\bibinfo{author}{\bibfnamefont{S.}~\bibnamefont{Sugawa}}, \bibnamefont{and}
	\bibinfo{author}{\bibfnamefont{I.~B.} \bibnamefont{Spielman}},
	\bibinfo{journal}{Phys. Rev. Lett.} \textbf{\bibinfo{volume}{124}},
	\bibinfo{pages}{053605} (\bibinfo{year}{2020}),
	\urlprefix\url{https://link.aps.org/doi/10.1103/PhysRevLett.124.053605}.
	
	\bibitem[{\citenamefont{Yao et~al.}(2022)\citenamefont{Yao, Zhang, and
			Chin}}]{Yao2022domain}
	\bibinfo{author}{\bibfnamefont{K.-X.} \bibnamefont{Yao}},
	\bibinfo{author}{\bibfnamefont{Z.}~\bibnamefont{Zhang}}, \bibnamefont{and}
	\bibinfo{author}{\bibfnamefont{C.}~\bibnamefont{Chin}},
	\bibinfo{journal}{Nature} \textbf{\bibinfo{volume}{602}}, \bibinfo{pages}{68}
	(\bibinfo{year}{2022}).
	
	\bibitem[{\citenamefont{Galitski and Spielman}(2013)}]{Galitski2013spin}
	\bibinfo{author}{\bibfnamefont{V.}~\bibnamefont{Galitski}} \bibnamefont{and}
	\bibinfo{author}{\bibfnamefont{I.~B.} \bibnamefont{Spielman}},
	\bibinfo{journal}{Nature} \textbf{\bibinfo{volume}{494}}, \bibinfo{pages}{49}
	(\bibinfo{year}{2013}).
	
	\bibitem[{\citenamefont{Zhai}(2015)}]{Zhai2015review}
	\bibinfo{author}{\bibfnamefont{H.}~\bibnamefont{Zhai}},
	\bibinfo{journal}{Reports on Progress in Physics}
	\textbf{\bibinfo{volume}{78}}, \bibinfo{pages}{026001}
	(\bibinfo{year}{2015}).
	
	\bibitem[{\citenamefont{Goldman et~al.}(2014)\citenamefont{Goldman,
			Juzeli{\=u}nas, {\"O}hberg, and Spielman}}]{Goldman2014light}
	\bibinfo{author}{\bibfnamefont{N.}~\bibnamefont{Goldman}},
	\bibinfo{author}{\bibfnamefont{G.}~\bibnamefont{Juzeli{\=u}nas}},
	\bibinfo{author}{\bibfnamefont{P.}~\bibnamefont{{\"O}hberg}},
	\bibnamefont{and} \bibinfo{author}{\bibfnamefont{I.~B.}
		\bibnamefont{Spielman}}, \bibinfo{journal}{Reports on Progress in Physics}
	\textbf{\bibinfo{volume}{77}}, \bibinfo{pages}{126401}
	(\bibinfo{year}{2014}).
	
	\bibitem[{\citenamefont{Atala et~al.}(2014)\citenamefont{Atala, Aidelsburger,
			Lohse, Barreiro, Paredes, and Bloch}}]{Atala2014observation}
	\bibinfo{author}{\bibfnamefont{M.}~\bibnamefont{Atala}},
	\bibinfo{author}{\bibfnamefont{M.}~\bibnamefont{Aidelsburger}},
	\bibinfo{author}{\bibfnamefont{M.}~\bibnamefont{Lohse}},
	\bibinfo{author}{\bibfnamefont{J.~T.} \bibnamefont{Barreiro}},
	\bibinfo{author}{\bibfnamefont{B.}~\bibnamefont{Paredes}}, \bibnamefont{and}
	\bibinfo{author}{\bibfnamefont{I.}~\bibnamefont{Bloch}},
	\bibinfo{journal}{Nature Physics} \textbf{\bibinfo{volume}{10}},
	\bibinfo{pages}{588} (\bibinfo{year}{2014}).
	
	\bibitem[{\citenamefont{Celi et~al.}(2014)\citenamefont{Celi, Massignan,
			Ruseckas, Goldman, Spielman, Juzeli\ifmmode~\bar{u}\else \={u}\fi{}nas, and
			Lewenstein}}]{Celi2014}
	\bibinfo{author}{\bibfnamefont{A.}~\bibnamefont{Celi}},
	\bibinfo{author}{\bibfnamefont{P.}~\bibnamefont{Massignan}},
	\bibinfo{author}{\bibfnamefont{J.}~\bibnamefont{Ruseckas}},
	\bibinfo{author}{\bibfnamefont{N.}~\bibnamefont{Goldman}},
	\bibinfo{author}{\bibfnamefont{I.~B.} \bibnamefont{Spielman}},
	\bibinfo{author}{\bibfnamefont{G.}~\bibnamefont{Juzeli\ifmmode~\bar{u}\else
			\={u}\fi{}nas}}, \bibnamefont{and}
	\bibinfo{author}{\bibfnamefont{M.}~\bibnamefont{Lewenstein}},
	\bibinfo{journal}{Phys. Rev. Lett.} \textbf{\bibinfo{volume}{112}},
	\bibinfo{pages}{043001} (\bibinfo{year}{2014}),
	\urlprefix\url{https://link.aps.org/doi/10.1103/PhysRevLett.112.043001}.
	
	\bibitem[{\citenamefont{Dhar et~al.}(2012)\citenamefont{Dhar, Maji, Mishra,
			Pai, Mukerjee, and Paramekanti}}]{Dhar2012}
	\bibinfo{author}{\bibfnamefont{A.}~\bibnamefont{Dhar}},
	\bibinfo{author}{\bibfnamefont{M.}~\bibnamefont{Maji}},
	\bibinfo{author}{\bibfnamefont{T.}~\bibnamefont{Mishra}},
	\bibinfo{author}{\bibfnamefont{R.~V.} \bibnamefont{Pai}},
	\bibinfo{author}{\bibfnamefont{S.}~\bibnamefont{Mukerjee}}, \bibnamefont{and}
	\bibinfo{author}{\bibfnamefont{A.}~\bibnamefont{Paramekanti}},
	\bibinfo{journal}{Phys. Rev. A} \textbf{\bibinfo{volume}{85}},
	\bibinfo{pages}{041602} (\bibinfo{year}{2012}),
	\urlprefix\url{https://link.aps.org/doi/10.1103/PhysRevA.85.041602}.
	
	\bibitem[{\citenamefont{Ho and Zhang}(2011)}]{Ho2011}
	\bibinfo{author}{\bibfnamefont{T.-L.} \bibnamefont{Ho}} \bibnamefont{and}
	\bibinfo{author}{\bibfnamefont{S.}~\bibnamefont{Zhang}},
	\bibinfo{journal}{Phys. Rev. Lett.} \textbf{\bibinfo{volume}{107}},
	\bibinfo{pages}{150403} (\bibinfo{year}{2011}),
	\urlprefix\url{https://link.aps.org/doi/10.1103/PhysRevLett.107.150403}.
	
	\bibitem[{\citenamefont{Li et~al.}(2012)\citenamefont{Li, Pitaevskii, and
			Stringari}}]{Li2012}
	\bibinfo{author}{\bibfnamefont{Y.}~\bibnamefont{Li}},
	\bibinfo{author}{\bibfnamefont{L.~P.} \bibnamefont{Pitaevskii}},
	\bibnamefont{and}
	\bibinfo{author}{\bibfnamefont{S.}~\bibnamefont{Stringari}},
	\bibinfo{journal}{Phys. Rev. Lett.} \textbf{\bibinfo{volume}{108}},
	\bibinfo{pages}{225301} (\bibinfo{year}{2012}),
	\urlprefix\url{https://link.aps.org/doi/10.1103/PhysRevLett.108.225301}.
	
	\bibitem[{\citenamefont{Hu et~al.}(2012)\citenamefont{Hu, Ramachandhran, Pu,
			and Liu}}]{Hu2012}
	\bibinfo{author}{\bibfnamefont{H.}~\bibnamefont{Hu}},
	\bibinfo{author}{\bibfnamefont{B.}~\bibnamefont{Ramachandhran}},
	\bibinfo{author}{\bibfnamefont{H.}~\bibnamefont{Pu}}, \bibnamefont{and}
	\bibinfo{author}{\bibfnamefont{X.-J.} \bibnamefont{Liu}},
	\bibinfo{journal}{Phys. Rev. Lett.} \textbf{\bibinfo{volume}{108}},
	\bibinfo{pages}{010402} (\bibinfo{year}{2012}),
	\urlprefix\url{https://link.aps.org/doi/10.1103/PhysRevLett.108.010402}.
	
	\bibitem[{\citenamefont{Cole et~al.}(2012)\citenamefont{Cole, Zhang,
			Paramekanti, and Trivedi}}]{Cole2012}
	\bibinfo{author}{\bibfnamefont{W.~S.} \bibnamefont{Cole}},
	\bibinfo{author}{\bibfnamefont{S.}~\bibnamefont{Zhang}},
	\bibinfo{author}{\bibfnamefont{A.}~\bibnamefont{Paramekanti}},
	\bibnamefont{and} \bibinfo{author}{\bibfnamefont{N.}~\bibnamefont{Trivedi}},
	\bibinfo{journal}{Phys. Rev. Lett.} \textbf{\bibinfo{volume}{109}},
	\bibinfo{pages}{085302} (\bibinfo{year}{2012}),
	\urlprefix\url{https://link.aps.org/doi/10.1103/PhysRevLett.109.085302}.
	
	\bibitem[{\citenamefont{Qu et~al.}(2013)\citenamefont{Qu, Hamner, Gong, Zhang,
			and Engels}}]{Qu2013}
	\bibinfo{author}{\bibfnamefont{C.}~\bibnamefont{Qu}},
	\bibinfo{author}{\bibfnamefont{C.}~\bibnamefont{Hamner}},
	\bibinfo{author}{\bibfnamefont{M.}~\bibnamefont{Gong}},
	\bibinfo{author}{\bibfnamefont{C.}~\bibnamefont{Zhang}}, \bibnamefont{and}
	\bibinfo{author}{\bibfnamefont{P.}~\bibnamefont{Engels}},
	\bibinfo{journal}{Phys. Rev. A} \textbf{\bibinfo{volume}{88}},
	\bibinfo{pages}{021604} (\bibinfo{year}{2013}),
	\urlprefix\url{https://link.aps.org/doi/10.1103/PhysRevA.88.021604}.
	
	\bibitem[{\citenamefont{Zheng et~al.}(2014)\citenamefont{Zheng, Liu, Miao,
			Chin, and Zhai}}]{Zheng2014}
	\bibinfo{author}{\bibfnamefont{W.}~\bibnamefont{Zheng}},
	\bibinfo{author}{\bibfnamefont{B.}~\bibnamefont{Liu}},
	\bibinfo{author}{\bibfnamefont{J.}~\bibnamefont{Miao}},
	\bibinfo{author}{\bibfnamefont{C.}~\bibnamefont{Chin}}, \bibnamefont{and}
	\bibinfo{author}{\bibfnamefont{H.}~\bibnamefont{Zhai}},
	\bibinfo{journal}{Phys. Rev. Lett.} \textbf{\bibinfo{volume}{113}},
	\bibinfo{pages}{155303} (\bibinfo{year}{2014}),
	\urlprefix\url{https://link.aps.org/doi/10.1103/PhysRevLett.113.155303}.
	
	\bibitem[{\citenamefont{Xu et~al.}(2014)\citenamefont{Xu, Cole, and
			Zhang}}]{Xu2014}
	\bibinfo{author}{\bibfnamefont{Z.}~\bibnamefont{Xu}},
	\bibinfo{author}{\bibfnamefont{W.~S.} \bibnamefont{Cole}}, \bibnamefont{and}
	\bibinfo{author}{\bibfnamefont{S.}~\bibnamefont{Zhang}},
	\bibinfo{journal}{Phys. Rev. A} \textbf{\bibinfo{volume}{89}},
	\bibinfo{pages}{051604} (\bibinfo{year}{2014}),
	\urlprefix\url{https://link.aps.org/doi/10.1103/PhysRevA.89.051604}.
	
	\bibitem[{\citenamefont{Khamehchi et~al.}(2016)\citenamefont{Khamehchi, Qu,
			Mossman, Zhang, and Engels}}]{Khamehchi2016spin}
	\bibinfo{author}{\bibfnamefont{M.}~\bibnamefont{Khamehchi}},
	\bibinfo{author}{\bibfnamefont{C.}~\bibnamefont{Qu}},
	\bibinfo{author}{\bibfnamefont{M.}~\bibnamefont{Mossman}},
	\bibinfo{author}{\bibfnamefont{C.}~\bibnamefont{Zhang}}, \bibnamefont{and}
	\bibinfo{author}{\bibfnamefont{P.}~\bibnamefont{Engels}},
	\bibinfo{journal}{Nature communications} \textbf{\bibinfo{volume}{7}},
	\bibinfo{pages}{10867} (\bibinfo{year}{2016}).
	
	\bibitem[{\citenamefont{Liu et~al.}(2016)\citenamefont{Liu, Clark, and
			Chin}}]{Liu2016}
	\bibinfo{author}{\bibfnamefont{T.}~\bibnamefont{Liu}},
	\bibinfo{author}{\bibfnamefont{L.~W.} \bibnamefont{Clark}}, \bibnamefont{and}
	\bibinfo{author}{\bibfnamefont{C.}~\bibnamefont{Chin}},
	\bibinfo{journal}{Phys. Rev. A} \textbf{\bibinfo{volume}{94}},
	\bibinfo{pages}{063646} (\bibinfo{year}{2016}),
	\urlprefix\url{https://link.aps.org/doi/10.1103/PhysRevA.94.063646}.
	
	\bibitem[{\citenamefont{Cole et~al.}(2019)\citenamefont{Cole, Lee, Mahmud,
			Alavirad, Spielman, and Sau}}]{Cole2019emergent}
	\bibinfo{author}{\bibfnamefont{W.~S.} \bibnamefont{Cole}},
	\bibinfo{author}{\bibfnamefont{J.}~\bibnamefont{Lee}},
	\bibinfo{author}{\bibfnamefont{K.~W.} \bibnamefont{Mahmud}},
	\bibinfo{author}{\bibfnamefont{Y.}~\bibnamefont{Alavirad}},
	\bibinfo{author}{\bibfnamefont{I.}~\bibnamefont{Spielman}}, \bibnamefont{and}
	\bibinfo{author}{\bibfnamefont{J.~D.} \bibnamefont{Sau}},
	\bibinfo{journal}{Scientific Reports} \textbf{\bibinfo{volume}{9}},
	\bibinfo{pages}{7471} (\bibinfo{year}{2019}).
	
	\bibitem[{\citenamefont{Orignac et~al.}(2017)\citenamefont{Orignac, Citro,
			Di~Dio, and De~Palo}}]{Orignac2017}
	\bibinfo{author}{\bibfnamefont{E.}~\bibnamefont{Orignac}},
	\bibinfo{author}{\bibfnamefont{R.}~\bibnamefont{Citro}},
	\bibinfo{author}{\bibfnamefont{M.}~\bibnamefont{Di~Dio}}, \bibnamefont{and}
	\bibinfo{author}{\bibfnamefont{S.}~\bibnamefont{De~Palo}},
	\bibinfo{journal}{Phys. Rev. B} \textbf{\bibinfo{volume}{96}},
	\bibinfo{pages}{014518} (\bibinfo{year}{2017}),
	\urlprefix\url{https://link.aps.org/doi/10.1103/PhysRevB.96.014518}.
	
	\bibitem[{\citenamefont{Tokuno and Georges}(2014)}]{Tokuno2014ground}
	\bibinfo{author}{\bibfnamefont{A.}~\bibnamefont{Tokuno}} \bibnamefont{and}
	\bibinfo{author}{\bibfnamefont{A.}~\bibnamefont{Georges}},
	\bibinfo{journal}{New Journal of Physics} \textbf{\bibinfo{volume}{16}},
	\bibinfo{pages}{073005} (\bibinfo{year}{2014}).
	
	\bibitem[{\citenamefont{Po et~al.}(2014)\citenamefont{Po, Chen, and
			Zhou}}]{Po2014}
	\bibinfo{author}{\bibfnamefont{H.~C.} \bibnamefont{Po}},
	\bibinfo{author}{\bibfnamefont{W.}~\bibnamefont{Chen}}, \bibnamefont{and}
	\bibinfo{author}{\bibfnamefont{Q.}~\bibnamefont{Zhou}},
	\bibinfo{journal}{Phys. Rev. A} \textbf{\bibinfo{volume}{90}},
	\bibinfo{pages}{011602} (\bibinfo{year}{2014}),
	\urlprefix\url{https://link.aps.org/doi/10.1103/PhysRevA.90.011602}.
	
	\bibitem[{\citenamefont{Po and Zhou}(2015)}]{Po2015}
	\bibinfo{author}{\bibfnamefont{H.~C.} \bibnamefont{Po}} \bibnamefont{and}
	\bibinfo{author}{\bibfnamefont{Q.}~\bibnamefont{Zhou}},
	\bibinfo{journal}{Nature Communications} \textbf{\bibinfo{volume}{6}},
	\bibinfo{pages}{8012} (\bibinfo{year}{2015}).
	
	\bibitem[{\citenamefont{Radi\ifmmode~\acute{c}\else \'{c}\fi{}
			et~al.}(2015)\citenamefont{Radi\ifmmode~\acute{c}\else \'{c}\fi{}, Natu, and
			Galitski}}]{Radic2015}
	\bibinfo{author}{\bibfnamefont{J.}~\bibnamefont{Radi\ifmmode~\acute{c}\else
			\'{c}\fi{}}}, \bibinfo{author}{\bibfnamefont{S.~S.} \bibnamefont{Natu}},
	\bibnamefont{and} \bibinfo{author}{\bibfnamefont{V.}~\bibnamefont{Galitski}},
	\bibinfo{journal}{Phys. Rev. A} \textbf{\bibinfo{volume}{91}},
	\bibinfo{pages}{063634} (\bibinfo{year}{2015}),
	\urlprefix\url{https://link.aps.org/doi/10.1103/PhysRevA.91.063634}.
	
	\bibitem[{\citenamefont{Sur and Yang}(2019)}]{Sur2019}
	\bibinfo{author}{\bibfnamefont{S.}~\bibnamefont{Sur}} \bibnamefont{and}
	\bibinfo{author}{\bibfnamefont{K.}~\bibnamefont{Yang}},
	\bibinfo{journal}{Phys. Rev. B} \textbf{\bibinfo{volume}{100}},
	\bibinfo{pages}{024519} (\bibinfo{year}{2019}),
	\urlprefix\url{https://link.aps.org/doi/10.1103/PhysRevB.100.024519}.
	
	\bibitem[{\citenamefont{Lake et~al.}(2021)\citenamefont{Lake, Senthil, and
			Vishwanath}}]{Lake2021}
	\bibinfo{author}{\bibfnamefont{E.}~\bibnamefont{Lake}},
	\bibinfo{author}{\bibfnamefont{T.}~\bibnamefont{Senthil}}, \bibnamefont{and}
	\bibinfo{author}{\bibfnamefont{A.}~\bibnamefont{Vishwanath}},
	\bibinfo{journal}{Phys. Rev. B} \textbf{\bibinfo{volume}{104}},
	\bibinfo{pages}{014517} (\bibinfo{year}{2021}),
	\urlprefix\url{https://link.aps.org/doi/10.1103/PhysRevB.104.014517}.
	
	\bibitem[{\citenamefont{Yang}(2004)}]{Yang2004}
	\bibinfo{author}{\bibfnamefont{K.}~\bibnamefont{Yang}}, \bibinfo{journal}{Phys.
		Rev. Lett.} \textbf{\bibinfo{volume}{93}}, \bibinfo{pages}{066401}
	(\bibinfo{year}{2004}),
	\urlprefix\url{https://link.aps.org/doi/10.1103/PhysRevLett.93.066401}.
	
	\bibitem[{\citenamefont{Sachdev and Senthil}(1996)}]{Sachdev1996}
	\bibinfo{author}{\bibfnamefont{S.}~\bibnamefont{Sachdev}} \bibnamefont{and}
	\bibinfo{author}{\bibfnamefont{T.}~\bibnamefont{Senthil}},
	\bibinfo{journal}{Annals of Physics} \textbf{\bibinfo{volume}{251}},
	\bibinfo{pages}{76} (\bibinfo{year}{1996}).
	
	\bibitem[{\citenamefont{Chamon}(2005)}]{Chamon2005}
	\bibinfo{author}{\bibfnamefont{C.}~\bibnamefont{Chamon}},
	\bibinfo{journal}{Phys. Rev. Lett.} \textbf{\bibinfo{volume}{94}},
	\bibinfo{pages}{040402} (\bibinfo{year}{2005}),
	\urlprefix\url{https://link.aps.org/doi/10.1103/PhysRevLett.94.040402}.
	
	\bibitem[{\citenamefont{Castelnovo and
			Chamon}(2012)}]{Castelnovo2012topological}
	\bibinfo{author}{\bibfnamefont{C.}~\bibnamefont{Castelnovo}} \bibnamefont{and}
	\bibinfo{author}{\bibfnamefont{C.}~\bibnamefont{Chamon}},
	\bibinfo{journal}{Philosophical Magazine} \textbf{\bibinfo{volume}{92}},
	\bibinfo{pages}{304} (\bibinfo{year}{2012}).
	
	\bibitem[{\citenamefont{Haah}(2011)}]{Haah2011}
	\bibinfo{author}{\bibfnamefont{J.}~\bibnamefont{Haah}}, \bibinfo{journal}{Phys.
		Rev. A} \textbf{\bibinfo{volume}{83}}, \bibinfo{pages}{042330}
	(\bibinfo{year}{2011}),
	\urlprefix\url{https://link.aps.org/doi/10.1103/PhysRevA.83.042330}.
	
	\bibitem[{\citenamefont{Vijay et~al.}(2016)\citenamefont{Vijay, Haah, and
			Fu}}]{Vijay2016}
	\bibinfo{author}{\bibfnamefont{S.}~\bibnamefont{Vijay}},
	\bibinfo{author}{\bibfnamefont{J.}~\bibnamefont{Haah}}, \bibnamefont{and}
	\bibinfo{author}{\bibfnamefont{L.}~\bibnamefont{Fu}}, \bibinfo{journal}{Phys.
		Rev. B} \textbf{\bibinfo{volume}{94}}, \bibinfo{pages}{235157}
	(\bibinfo{year}{2016}),
	\urlprefix\url{https://link.aps.org/doi/10.1103/PhysRevB.94.235157}.
	
	\bibitem[{\citenamefont{Pretko}(2017{\natexlab{a}})}]{Pretko2017}
	\bibinfo{author}{\bibfnamefont{M.}~\bibnamefont{Pretko}},
	\bibinfo{journal}{Phys. Rev. B} \textbf{\bibinfo{volume}{95}},
	\bibinfo{pages}{115139} (\bibinfo{year}{2017}{\natexlab{a}}),
	\urlprefix\url{https://link.aps.org/doi/10.1103/PhysRevB.95.115139}.
	
	\bibitem[{\citenamefont{Pretko}(2017{\natexlab{b}})}]{Pretko2017Generalized}
	\bibinfo{author}{\bibfnamefont{M.}~\bibnamefont{Pretko}},
	\bibinfo{journal}{Phys. Rev. B} \textbf{\bibinfo{volume}{96}},
	\bibinfo{pages}{035119} (\bibinfo{year}{2017}{\natexlab{b}}),
	\urlprefix\url{https://link.aps.org/doi/10.1103/PhysRevB.96.035119}.
	
	\bibitem[{\citenamefont{Prem et~al.}(2017)\citenamefont{Prem, Haah, and
			Nandkishore}}]{Prem2017}
	\bibinfo{author}{\bibfnamefont{A.}~\bibnamefont{Prem}},
	\bibinfo{author}{\bibfnamefont{J.}~\bibnamefont{Haah}}, \bibnamefont{and}
	\bibinfo{author}{\bibfnamefont{R.}~\bibnamefont{Nandkishore}},
	\bibinfo{journal}{Phys. Rev. B} \textbf{\bibinfo{volume}{95}},
	\bibinfo{pages}{155133} (\bibinfo{year}{2017}),
	\urlprefix\url{https://link.aps.org/doi/10.1103/PhysRevB.95.155133}.
	
	\bibitem[{\citenamefont{Nandkishore and Hermele}(2019)}]{FractonsReview}
	\bibinfo{author}{\bibfnamefont{R.~M.} \bibnamefont{Nandkishore}}
	\bibnamefont{and} \bibinfo{author}{\bibfnamefont{M.}~\bibnamefont{Hermele}},
	\bibinfo{journal}{Annual Review of Condensed Matter Physics}
	\textbf{\bibinfo{volume}{10}}, \bibinfo{pages}{295} (\bibinfo{year}{2019}),
	\urlprefix\url{https://doi.org/10.1146/annurev-conmatphys-031218-013604}.
	
	\bibitem[{\citenamefont{Gromov et~al.}(2020)\citenamefont{Gromov, Lucas, and
			Nandkishore}}]{Gromov2020}
	\bibinfo{author}{\bibfnamefont{A.}~\bibnamefont{Gromov}},
	\bibinfo{author}{\bibfnamefont{A.}~\bibnamefont{Lucas}}, \bibnamefont{and}
	\bibinfo{author}{\bibfnamefont{R.~M.} \bibnamefont{Nandkishore}},
	\bibinfo{journal}{Phys. Rev. Res.} \textbf{\bibinfo{volume}{2}},
	\bibinfo{pages}{033124} (\bibinfo{year}{2020}),
	\urlprefix\url{https://link.aps.org/doi/10.1103/PhysRevResearch.2.033124}.
	
	\bibitem[{\citenamefont{Pretko et~al.}(2020)\citenamefont{Pretko, Chen, and
			You}}]{Pretko2020fracton}
	\bibinfo{author}{\bibfnamefont{M.}~\bibnamefont{Pretko}},
	\bibinfo{author}{\bibfnamefont{X.}~\bibnamefont{Chen}}, \bibnamefont{and}
	\bibinfo{author}{\bibfnamefont{Y.}~\bibnamefont{You}},
	\bibinfo{journal}{International Journal of Modern Physics A}
	\textbf{\bibinfo{volume}{35}}, \bibinfo{pages}{2030003}
	(\bibinfo{year}{2020}).
	
	\bibitem[{\citenamefont{Radzihovsky}(2020)}]{Radzihovsky2020}
	\bibinfo{author}{\bibfnamefont{L.}~\bibnamefont{Radzihovsky}},
	\bibinfo{journal}{Phys. Rev. Lett.} \textbf{\bibinfo{volume}{125}},
	\bibinfo{pages}{267601} (\bibinfo{year}{2020}),
	\urlprefix\url{https://link.aps.org/doi/10.1103/PhysRevLett.125.267601}.
	
	\bibitem[{\citenamefont{Gromov and Radzihovsky}(2024)}]{Gromov2022fracton}
	\bibinfo{author}{\bibfnamefont{A.}~\bibnamefont{Gromov}} \bibnamefont{and}
	\bibinfo{author}{\bibfnamefont{L.}~\bibnamefont{Radzihovsky}},
	\bibinfo{journal}{Rev. Mod. Phys.} \textbf{\bibinfo{volume}{96}},
	\bibinfo{pages}{011001} (\bibinfo{year}{2024}),
	\urlprefix\url{https://link.aps.org/doi/10.1103/RevModPhys.96.011001}.
	
	\bibitem[{\citenamefont{Seiberg}(2020)}]{Seiberg2020field}
	\bibinfo{author}{\bibfnamefont{N.}~\bibnamefont{Seiberg}},
	\bibinfo{journal}{SciPost Physics} \textbf{\bibinfo{volume}{8}},
	\bibinfo{pages}{050} (\bibinfo{year}{2020}).
	
	\bibitem[{\citenamefont{Pai and Pretko}(2020)}]{Pai2020}
	\bibinfo{author}{\bibfnamefont{S.}~\bibnamefont{Pai}} \bibnamefont{and}
	\bibinfo{author}{\bibfnamefont{M.}~\bibnamefont{Pretko}},
	\bibinfo{journal}{Phys. Rev. Res.} \textbf{\bibinfo{volume}{2}},
	\bibinfo{pages}{013094} (\bibinfo{year}{2020}),
	\urlprefix\url{https://link.aps.org/doi/10.1103/PhysRevResearch.2.013094}.
	
	\bibitem[{\citenamefont{Lake et~al.}(2022)\citenamefont{Lake, Hermele, and
			Senthil}}]{Lake2022}
	\bibinfo{author}{\bibfnamefont{E.}~\bibnamefont{Lake}},
	\bibinfo{author}{\bibfnamefont{M.}~\bibnamefont{Hermele}}, \bibnamefont{and}
	\bibinfo{author}{\bibfnamefont{T.}~\bibnamefont{Senthil}},
	\bibinfo{journal}{Phys. Rev. B} \textbf{\bibinfo{volume}{106}},
	\bibinfo{pages}{064511} (\bibinfo{year}{2022}),
	\urlprefix\url{https://link.aps.org/doi/10.1103/PhysRevB.106.064511}.
	
	\bibitem[{\citenamefont{Gorantla
			et~al.}(2022{\natexlab{a}})\citenamefont{Gorantla, Lam, Seiberg, and
			Shao}}]{Gorantla2022}
	\bibinfo{author}{\bibfnamefont{P.}~\bibnamefont{Gorantla}},
	\bibinfo{author}{\bibfnamefont{H.~T.} \bibnamefont{Lam}},
	\bibinfo{author}{\bibfnamefont{N.}~\bibnamefont{Seiberg}}, \bibnamefont{and}
	\bibinfo{author}{\bibfnamefont{S.-H.} \bibnamefont{Shao}},
	\bibinfo{journal}{Phys. Rev. B} \textbf{\bibinfo{volume}{106}},
	\bibinfo{pages}{045112} (\bibinfo{year}{2022}{\natexlab{a}}),
	\urlprefix\url{https://link.aps.org/doi/10.1103/PhysRevB.106.045112}.
	
	\bibitem[{\citenamefont{Gorantla
			et~al.}(2022{\natexlab{b}})\citenamefont{Gorantla, Lam, Seiberg, and
			Shao}}]{Gorantla2022_2plus1}
	\bibinfo{author}{\bibfnamefont{P.}~\bibnamefont{Gorantla}},
	\bibinfo{author}{\bibfnamefont{H.~T.} \bibnamefont{Lam}},
	\bibinfo{author}{\bibfnamefont{N.}~\bibnamefont{Seiberg}}, \bibnamefont{and}
	\bibinfo{author}{\bibfnamefont{S.-H.} \bibnamefont{Shao}},
	\bibinfo{journal}{arXiv preprint arXiv:2209.10030}
	(\bibinfo{year}{2022}{\natexlab{b}}).
	
	\bibitem[{\citenamefont{Zechmann et~al.}(2023)\citenamefont{Zechmann, Altman,
			Knap, and Feldmeier}}]{Zechmann2022fractonic}
	\bibinfo{author}{\bibfnamefont{P.}~\bibnamefont{Zechmann}},
	\bibinfo{author}{\bibfnamefont{E.}~\bibnamefont{Altman}},
	\bibinfo{author}{\bibfnamefont{M.}~\bibnamefont{Knap}}, \bibnamefont{and}
	\bibinfo{author}{\bibfnamefont{J.}~\bibnamefont{Feldmeier}},
	\bibinfo{journal}{Phys. Rev. B} \textbf{\bibinfo{volume}{107}},
	\bibinfo{pages}{195131} (\bibinfo{year}{2023}),
	\urlprefix\url{https://link.aps.org/doi/10.1103/PhysRevB.107.195131}.
	
	\bibitem[{\citenamefont{Radzihovsky}(2022)}]{Radzihovsky2022}
	\bibinfo{author}{\bibfnamefont{L.}~\bibnamefont{Radzihovsky}},
	\bibinfo{journal}{Phys. Rev. B} \textbf{\bibinfo{volume}{106}},
	\bibinfo{pages}{224510} (\bibinfo{year}{2022}),
	\urlprefix\url{https://link.aps.org/doi/10.1103/PhysRevB.106.224510}.
	
	\bibitem[{\citenamefont{Lake and Senthil}(2023)}]{Lake2023}
	\bibinfo{author}{\bibfnamefont{E.}~\bibnamefont{Lake}} \bibnamefont{and}
	\bibinfo{author}{\bibfnamefont{T.}~\bibnamefont{Senthil}},
	\bibinfo{journal}{arXiv preprint arXiv:2302.08499}  (\bibinfo{year}{2023}).
	
	\bibitem[{\citenamefont{Sachdev and Young}(1997)}]{Sachdev1997}
	\bibinfo{author}{\bibfnamefont{S.}~\bibnamefont{Sachdev}} \bibnamefont{and}
	\bibinfo{author}{\bibfnamefont{A.~P.} \bibnamefont{Young}},
	\bibinfo{journal}{Phys. Rev. Lett.} \textbf{\bibinfo{volume}{78}},
	\bibinfo{pages}{2220} (\bibinfo{year}{1997}),
	\urlprefix\url{https://link.aps.org/doi/10.1103/PhysRevLett.78.2220}.
	
	\bibitem[{\citenamefont{Sachdev}(2011)}]{Sachdev1999quantum}
	\bibinfo{author}{\bibfnamefont{S.}~\bibnamefont{Sachdev}},
	\emph{\bibinfo{title}{Quantum Phase Transitions}}
	(\bibinfo{publisher}{Cambridge University Press}, \bibinfo{year}{2011}).
	
	\bibitem[{\citenamefont{Albuquerque et~al.}(2009)\citenamefont{Albuquerque,
			Hamer, and Oitmaa}}]{Albuquerque2009}
	\bibinfo{author}{\bibfnamefont{A.~F.} \bibnamefont{Albuquerque}},
	\bibinfo{author}{\bibfnamefont{C.~J.} \bibnamefont{Hamer}}, \bibnamefont{and}
	\bibinfo{author}{\bibfnamefont{J.}~\bibnamefont{Oitmaa}},
	\bibinfo{journal}{Phys. Rev. B} \textbf{\bibinfo{volume}{79}},
	\bibinfo{pages}{054412} (\bibinfo{year}{2009}),
	\urlprefix\url{https://link.aps.org/doi/10.1103/PhysRevB.79.054412}.
	
	\bibitem[{\citenamefont{Lieb and Liniger}(1963)}]{LiebLiniger}
	\bibinfo{author}{\bibfnamefont{E.~H.} \bibnamefont{Lieb}} \bibnamefont{and}
	\bibinfo{author}{\bibfnamefont{W.}~\bibnamefont{Liniger}},
	\bibinfo{journal}{Phys. Rev.} \textbf{\bibinfo{volume}{130}},
	\bibinfo{pages}{1605} (\bibinfo{year}{1963}),
	\urlprefix\url{https://link.aps.org/doi/10.1103/PhysRev.130.1605}.
	
	\bibitem[{\citenamefont{Haldane}(1981)}]{Haldane1981}
	\bibinfo{author}{\bibfnamefont{F.~D.~M.} \bibnamefont{Haldane}},
	\bibinfo{journal}{Phys. Rev. Lett.} \textbf{\bibinfo{volume}{47}},
	\bibinfo{pages}{1840} (\bibinfo{year}{1981}),
	\urlprefix\url{https://link.aps.org/doi/10.1103/PhysRevLett.47.1840}.
	
	\bibitem[{SM()}]{SM}
	\bibinfo{note}{See supplemental material, including
		Refs.~\cite{Sachdev1999quantum,Sachdev1997,Kamenev2009keldysh,Hoyos2013lifshitz,Hoyos2014lifshitz,Yang2004,LiebLiniger,Liu2016,Cole2019emergent},
		for derivations and extended discussions}.
	
	\bibitem[{\citenamefont{Preiss et~al.}(2015)\citenamefont{Preiss, Ma, Tai,
			Lukin, Rispoli, Zupancic, Lahini, Islam, and Greiner}}]{Preiss2015strongly}
	\bibinfo{author}{\bibfnamefont{P.~M.} \bibnamefont{Preiss}},
	\bibinfo{author}{\bibfnamefont{R.}~\bibnamefont{Ma}},
	\bibinfo{author}{\bibfnamefont{M.~E.} \bibnamefont{Tai}},
	\bibinfo{author}{\bibfnamefont{A.}~\bibnamefont{Lukin}},
	\bibinfo{author}{\bibfnamefont{M.}~\bibnamefont{Rispoli}},
	\bibinfo{author}{\bibfnamefont{P.}~\bibnamefont{Zupancic}},
	\bibinfo{author}{\bibfnamefont{Y.}~\bibnamefont{Lahini}},
	\bibinfo{author}{\bibfnamefont{R.}~\bibnamefont{Islam}}, \bibnamefont{and}
	\bibinfo{author}{\bibfnamefont{M.}~\bibnamefont{Greiner}},
	\bibinfo{journal}{Science} \textbf{\bibinfo{volume}{347}},
	\bibinfo{pages}{1229} (\bibinfo{year}{2015}).
	
	\bibitem[{\citenamefont{Yang et~al.}(2020)\citenamefont{Yang, Liu, Gorshkov,
			and Iadecola}}]{Yang2020}
	\bibinfo{author}{\bibfnamefont{Z.-C.} \bibnamefont{Yang}},
	\bibinfo{author}{\bibfnamefont{F.}~\bibnamefont{Liu}},
	\bibinfo{author}{\bibfnamefont{A.~V.} \bibnamefont{Gorshkov}},
	\bibnamefont{and} \bibinfo{author}{\bibfnamefont{T.}~\bibnamefont{Iadecola}},
	\bibinfo{journal}{Phys. Rev. Lett.} \textbf{\bibinfo{volume}{124}},
	\bibinfo{pages}{207602} (\bibinfo{year}{2020}),
	\urlprefix\url{https://link.aps.org/doi/10.1103/PhysRevLett.124.207602}.
	
	\bibitem[{\citenamefont{Bastianello et~al.}(2022)\citenamefont{Bastianello,
			Borla, and Moroz}}]{Bastianello2022}
	\bibinfo{author}{\bibfnamefont{A.}~\bibnamefont{Bastianello}},
	\bibinfo{author}{\bibfnamefont{U.}~\bibnamefont{Borla}}, \bibnamefont{and}
	\bibinfo{author}{\bibfnamefont{S.}~\bibnamefont{Moroz}},
	\bibinfo{journal}{Phys. Rev. Lett.} \textbf{\bibinfo{volume}{128}},
	\bibinfo{pages}{196601} (\bibinfo{year}{2022}),
	\urlprefix\url{https://link.aps.org/doi/10.1103/PhysRevLett.128.196601}.
	
	\bibitem[{\citenamefont{Khemani et~al.}(2020)\citenamefont{Khemani, Hermele,
			and Nandkishore}}]{Khemani2020}
	\bibinfo{author}{\bibfnamefont{V.}~\bibnamefont{Khemani}},
	\bibinfo{author}{\bibfnamefont{M.}~\bibnamefont{Hermele}}, \bibnamefont{and}
	\bibinfo{author}{\bibfnamefont{R.}~\bibnamefont{Nandkishore}},
	\bibinfo{journal}{Phys. Rev. B} \textbf{\bibinfo{volume}{101}},
	\bibinfo{pages}{174204} (\bibinfo{year}{2020}),
	\urlprefix\url{https://link.aps.org/doi/10.1103/PhysRevB.101.174204}.
	
	\bibitem[{\citenamefont{Sala et~al.}(2020)\citenamefont{Sala, Rakovszky,
			Verresen, Knap, and Pollmann}}]{Sala2020}
	\bibinfo{author}{\bibfnamefont{P.}~\bibnamefont{Sala}},
	\bibinfo{author}{\bibfnamefont{T.}~\bibnamefont{Rakovszky}},
	\bibinfo{author}{\bibfnamefont{R.}~\bibnamefont{Verresen}},
	\bibinfo{author}{\bibfnamefont{M.}~\bibnamefont{Knap}}, \bibnamefont{and}
	\bibinfo{author}{\bibfnamefont{F.}~\bibnamefont{Pollmann}},
	\bibinfo{journal}{Phys. Rev. X} \textbf{\bibinfo{volume}{10}},
	\bibinfo{pages}{011047} (\bibinfo{year}{2020}),
	\urlprefix\url{https://link.aps.org/doi/10.1103/PhysRevX.10.011047}.
	
	\bibitem[{\citenamefont{Rakovszky et~al.}(2020)\citenamefont{Rakovszky, Sala,
			Verresen, Knap, and Pollmann}}]{Rakovszky2020}
	\bibinfo{author}{\bibfnamefont{T.}~\bibnamefont{Rakovszky}},
	\bibinfo{author}{\bibfnamefont{P.}~\bibnamefont{Sala}},
	\bibinfo{author}{\bibfnamefont{R.}~\bibnamefont{Verresen}},
	\bibinfo{author}{\bibfnamefont{M.}~\bibnamefont{Knap}}, \bibnamefont{and}
	\bibinfo{author}{\bibfnamefont{F.}~\bibnamefont{Pollmann}},
	\bibinfo{journal}{Phys. Rev. B} \textbf{\bibinfo{volume}{101}},
	\bibinfo{pages}{125126} (\bibinfo{year}{2020}),
	\urlprefix\url{https://link.aps.org/doi/10.1103/PhysRevB.101.125126}.
	
	\bibitem[{\citenamefont{De~Tomasi et~al.}(2019)\citenamefont{De~Tomasi,
			Hetterich, Sala, and Pollmann}}]{DeTomasi2019}
	\bibinfo{author}{\bibfnamefont{G.}~\bibnamefont{De~Tomasi}},
	\bibinfo{author}{\bibfnamefont{D.}~\bibnamefont{Hetterich}},
	\bibinfo{author}{\bibfnamefont{P.}~\bibnamefont{Sala}}, \bibnamefont{and}
	\bibinfo{author}{\bibfnamefont{F.}~\bibnamefont{Pollmann}},
	\bibinfo{journal}{Phys. Rev. B} \textbf{\bibinfo{volume}{100}},
	\bibinfo{pages}{214313} (\bibinfo{year}{2019}),
	\urlprefix\url{https://link.aps.org/doi/10.1103/PhysRevB.100.214313}.
	
	\bibitem[{\citenamefont{Moudgalya and Motrunich}(2022)}]{Moudgalya2022}
	\bibinfo{author}{\bibfnamefont{S.}~\bibnamefont{Moudgalya}} \bibnamefont{and}
	\bibinfo{author}{\bibfnamefont{O.~I.} \bibnamefont{Motrunich}},
	\bibinfo{journal}{Phys. Rev. X} \textbf{\bibinfo{volume}{12}},
	\bibinfo{pages}{011050} (\bibinfo{year}{2022}),
	\urlprefix\url{https://link.aps.org/doi/10.1103/PhysRevX.12.011050}.
	
	\bibitem[{\citenamefont{Kohlert et~al.}(2021)\citenamefont{Kohlert, Scherg,
			Sala, Pollmann, Madhusudhana, Bloch, and
			Aidelsburger}}]{Kohlert2021experimental}
	\bibinfo{author}{\bibfnamefont{T.}~\bibnamefont{Kohlert}},
	\bibinfo{author}{\bibfnamefont{S.}~\bibnamefont{Scherg}},
	\bibinfo{author}{\bibfnamefont{P.}~\bibnamefont{Sala}},
	\bibinfo{author}{\bibfnamefont{F.}~\bibnamefont{Pollmann}},
	\bibinfo{author}{\bibfnamefont{B.~H.} \bibnamefont{Madhusudhana}},
	\bibinfo{author}{\bibfnamefont{I.}~\bibnamefont{Bloch}}, \bibnamefont{and}
	\bibinfo{author}{\bibfnamefont{M.}~\bibnamefont{Aidelsburger}},
	\bibinfo{journal}{arXiv preprint arXiv:2106.15586}  (\bibinfo{year}{2021}).
	
	\bibitem[{\citenamefont{Mukherjee et~al.}(2021)\citenamefont{Mukherjee,
			Banerjee, Sengupta, and Sen}}]{Mukherjee2021}
	\bibinfo{author}{\bibfnamefont{B.}~\bibnamefont{Mukherjee}},
	\bibinfo{author}{\bibfnamefont{D.}~\bibnamefont{Banerjee}},
	\bibinfo{author}{\bibfnamefont{K.}~\bibnamefont{Sengupta}}, \bibnamefont{and}
	\bibinfo{author}{\bibfnamefont{A.}~\bibnamefont{Sen}},
	\bibinfo{journal}{Phys. Rev. B} \textbf{\bibinfo{volume}{104}},
	\bibinfo{pages}{155117} (\bibinfo{year}{2021}),
	\urlprefix\url{https://link.aps.org/doi/10.1103/PhysRevB.104.155117}.
	
	\bibitem[{\citenamefont{Ghosh et~al.}(2023)\citenamefont{Ghosh, Paul, and
			Sengupta}}]{Ghosh2023prethermal}
	\bibinfo{author}{\bibfnamefont{S.}~\bibnamefont{Ghosh}},
	\bibinfo{author}{\bibfnamefont{I.}~\bibnamefont{Paul}}, \bibnamefont{and}
	\bibinfo{author}{\bibfnamefont{K.}~\bibnamefont{Sengupta}},
	\bibinfo{journal}{Physical Review Letters} \textbf{\bibinfo{volume}{130}},
	\bibinfo{pages}{120401} (\bibinfo{year}{2023}).
	
	\bibitem[{\citenamefont{Lan et~al.}(2018)\citenamefont{Lan, van Horssen,
			Powell, and Garrahan}}]{Lan2018}
	\bibinfo{author}{\bibfnamefont{Z.}~\bibnamefont{Lan}},
	\bibinfo{author}{\bibfnamefont{M.}~\bibnamefont{van Horssen}},
	\bibinfo{author}{\bibfnamefont{S.}~\bibnamefont{Powell}}, \bibnamefont{and}
	\bibinfo{author}{\bibfnamefont{J.~P.} \bibnamefont{Garrahan}},
	\bibinfo{journal}{Phys. Rev. Lett.} \textbf{\bibinfo{volume}{121}},
	\bibinfo{pages}{040603} (\bibinfo{year}{2018}),
	\urlprefix\url{https://link.aps.org/doi/10.1103/PhysRevLett.121.040603}.
	
	\bibitem[{\citenamefont{Kozii et~al.}(2017)\citenamefont{Kozii, Ruhman, Fu, and
			Radzihovsky}}]{Kozii2017}
	\bibinfo{author}{\bibfnamefont{V.}~\bibnamefont{Kozii}},
	\bibinfo{author}{\bibfnamefont{J.}~\bibnamefont{Ruhman}},
	\bibinfo{author}{\bibfnamefont{L.}~\bibnamefont{Fu}}, \bibnamefont{and}
	\bibinfo{author}{\bibfnamefont{L.}~\bibnamefont{Radzihovsky}},
	\bibinfo{journal}{Phys. Rev. B} \textbf{\bibinfo{volume}{96}},
	\bibinfo{pages}{094419} (\bibinfo{year}{2017}),
	\urlprefix\url{https://link.aps.org/doi/10.1103/PhysRevB.96.094419}.
	
	\bibitem[{\citenamefont{Kamenev and Levchenko}(2009)}]{Kamenev2009keldysh}
	\bibinfo{author}{\bibfnamefont{A.}~\bibnamefont{Kamenev}} \bibnamefont{and}
	\bibinfo{author}{\bibfnamefont{A.}~\bibnamefont{Levchenko}},
	\bibinfo{journal}{Advances in Physics} \textbf{\bibinfo{volume}{58}},
	\bibinfo{pages}{197} (\bibinfo{year}{2009}).
	
	\bibitem[{\citenamefont{Hoyos et~al.}(2013)\citenamefont{Hoyos, Kim, and
			Oz}}]{Hoyos2013lifshitz}
	\bibinfo{author}{\bibfnamefont{C.}~\bibnamefont{Hoyos}},
	\bibinfo{author}{\bibfnamefont{B.~S.} \bibnamefont{Kim}}, \bibnamefont{and}
	\bibinfo{author}{\bibfnamefont{Y.}~\bibnamefont{Oz}},
	\bibinfo{journal}{Journal of High Energy Physics}
	\textbf{\bibinfo{volume}{2013}}, \bibinfo{pages}{1} (\bibinfo{year}{2013}).
	
	\bibitem[{\citenamefont{Hoyos et~al.}(2014)\citenamefont{Hoyos, Kim, and
			Oz}}]{Hoyos2014lifshitz}
	\bibinfo{author}{\bibfnamefont{C.}~\bibnamefont{Hoyos}},
	\bibinfo{author}{\bibfnamefont{B.~S.} \bibnamefont{Kim}}, \bibnamefont{and}
	\bibinfo{author}{\bibfnamefont{Y.}~\bibnamefont{Oz}},
	\bibinfo{journal}{Journal of High Energy Physics}
	\textbf{\bibinfo{volume}{2014}}, \bibinfo{pages}{1} (\bibinfo{year}{2014}).
	
\end{thebibliography}
\end{document}